\documentclass{nature}

\usepackage[numbers,super,sort,compress]{natbib}

\usepackage{times}

\usepackage{graphicx}
\usepackage{bm}
\usepackage{color}
\usepackage[normalem]{ulem}
\usepackage{amsmath}
\usepackage{amssymb}
\usepackage{hyperref}
\hypersetup{colorlinks=false, linkbordercolor={0 0 1}, citecolor=blue, urlcolor=blue, linkcolor=blue, }

\usepackage{cleveref}
\usepackage{bookmark}
\usepackage{caption}
\usepackage{xcolor}

\newcommand{\angstrom}{\mbox{\normalfont\AA}}

\newcommand\redsout{\bgroup\markoverwith{\textcolor{red}{\rule[0.5ex]{2pt}{0.4pt}}}\ULon}

\newcommand{\nutotal}{$\nu_{total}$}	
\newcommand{\red}{\textcolor{red}}
\newcommand{\blue}{\textcolor{blue}}
\newcommand{\green}{\textcolor{green}}
\newcommand{\CF}{$_{1}^{2}$CF }
\newcommand{\CFs}{$_{1}^{2}$CFs }
\newcommand{\dlb}{$d/\ell_B$}
\newcommand{\up}{$|\uparrow >$}
\newcommand{\down}{$|\downarrow >$}
\newcommand{\ttCF}{$_{2}^{2}$CF }
\newcommand{\ttCFs}{$_{2}^{2}$CFs }
\newcommand{\tzCF}{$_{0}^{2}$CF }
\newcommand{\tzCFs}{$_{0}^{2}$CFs }
\newcommand{\LLL}{$\Lambda$-level }
\newcommand{\RCF}{$R_{xx}^{CF}$ }
\newcommand{\dragxy}{$R_{xy}^{drag}$ }
\newcommand{\dragxx}{$R_{xx}^{drag}$ }

\makeatletter
\let\saved@includegraphics\includegraphics
\AtBeginDocument{\let\includegraphics\saved@includegraphics}
\makeatother

\title{Real-space Imaging of Quantum Hall Quasiparticles} 

\author
{Jinghao Deng$^{1\dagger}$, Yiming Sun$^{1, 2\dagger}$, Dimitri Pimenov$^{1}$, Takashi Taniguchi$^{3}$, \\ Kenji Watanabe$^{4}$, Erich J Mueller$^{1\ast}$, Xiaomeng Liu$^{1\ast}$\\
\\
\normalsize{$^{1}$Laboratory of Atomic and Solid State Physics, Cornell University,}\\ 
\normalsize{Ithaca, NY, USA}\\
\normalsize{$^{2}$School of Applied and Engineering Physics, Cornell University,}\\
\normalsize{Ithaca, NY, USA}\\
\normalsize{$^{3}$Research Center for Functional Materials, National Institute for Materials Science,}\\
\normalsize{Tsukuba, Japan}\\
\normalsize{$^{4}$Research Center for Electronic and Optical Materials, National Institute for Materials Science,}\\
\normalsize{Tsukuba, Japan}\\
\normalsize{$\dagger$ These authors contributed equally: Jinghao Deng, Yiming Sun}\\
\\
\normalsize{$^\ast$To whom correspondence should be addressed:}\\
\normalsize{X.L. (email: xl956@cornell.edu); E.J.M. (email: em256@cornell.edu)}
}

\date{}

\begin{document} 


\baselineskip24pt


\maketitle 

\pagebreak

\phantomsection
\addcontentsline{toc}{section}{Abstract}
\begin{abstract}
Quantum Hall systems host emergent quasiparticles with unusual charge, spin, and statistics, such as fractionally charged anyons \cite{Laughlin1983AnomalousQuantumHall, Halperin1984StatisticsQuasiparticlesHierarchy}. Although transport measurements have revealed many of their collective properties\cite{Saminadayar1997Observation$mathitemathit3$, de-Picciotto1997Directobservationfractional, Nakamura2020Directobservationanyonic, Bartolomei2020Fractionalstatisticsanyon}, identifying and visualizing individual quasiparticles remain elusive. Here we use scanning tunneling spectroscopy (STS) to image quantum Hall quasiparticles in graphene. Within incompressible quantum Hall states, we observe spatial variation of Landau level energies originating from electrostatic potentials created by charged defects in graphene and the underlying hexagonal boron nitride (hBN). For surface and near-surface defects, the Coulomb potential lifts the degeneracy of Landau orbitals, producing discrete energy splittings that reveal Landau orbital wavefunctions. In quantum Hall ferromagnetic states, quasiparticles bound to defect potentials produce distinct spatial and spectroscopic signatures that serve as hallmarks of the presence and number of localized excitations. In the fractional quantum Hall regime at one-third filling, our theoretical calculations predict discrete spectroscopic changes associated with the sequential addition of localized anyons, with a three-anyon bound state quantitatively reproducing our experimental data at $\nu = 5/3$. These observations establish spectroscopic fingerprints of quantum Hall quasiparticles and provide a pathway toward imaging and manipulating individual anyons in real space.
\end{abstract}

\phantomsection
\addcontentsline{toc}{section}{Main Text}

The quantum Hall effect represents one of the clearest examples of emergent phenomena in condensed matter physics. Under strong magnetic fields, interacting electrons in two dimensions organize into topologically ordered states that host quasiparticles with properties fundamentally different from those of the constituent electrons. In spin or valley polarized quantum Hall ferromagnetic states, the elementary excitation is expected to be a topological spin texture that carries an electric charge, known as a skyrmion \cite{Sondhi1993Skyrmionscrossoverinteger,Schmeller1995EvidenceSkyrmionsSingle,MacDonald1996SkyrmionsSigmaModels,Yang2006Collectivemodesskyrmion,Lian2017Spinvalleyskyrmionsgraphene, liuVisualizingBrokenSymmetry2022, Halperin2020FractionalQuantumHall, ezawa2013quantum}. When the Landau level (LL) is instead filled to a fractional number of electrons per orbital, the system can form fractional quantum Hall states that host fractionalized quasiparticles known as anyons that carry fractional charges and obey exotic exchange statistics \cite{Wilczek1982MagneticFluxAngular,Wilczek1982QuantumMechanicsFractionalSpin,Tsui1982TwoDimensionalMagnetotransportExtreme,Laughlin1983AnomalousQuantumHall,Halperin1984StatisticsQuasiparticlesHierarchy,Jain1990Theoryfractionalquantum,Eisenstein1990FractionalQuantumHall,Du1993Experimentalevidencenew, Halperin2020FractionalQuantumHall}. Furthermore, non-Abelian anyons, theorized in even denominator fractional quantum Hall states \cite{Willett1987Observationevendenominatorquantum,Moore1991Nonabelionsfractionalquantum,Nayak2008NonAbeliananyonstopological}, can encode information in their topologically protected quantum states and are central to proposals for fault-tolerant quantum computation.

Experimental evidence for these quasiparticles has largely come from bulk transport measurements. Fractional charge has been detected through shot-noise experiments \cite{Saminadayar1997Observation$mathitemathit3$,de-Picciotto1997Directobservationfractional}, and recent interferometry studies have provided signatures consistent with Abelian anyonic statistics \cite{Nakamura2020Directobservationanyonic, 
Werkmeister2025,henzinger2026controlledlocalizationanyonsgraphene,
fwjg-mx9h}. Despite these advances, transport measurements probe quasiparticles only indirectly through their interaction with edge states \cite{Kim2026AharonovBohminterference} and offer limited insight into their microscopic structure. Scanning single-electron-transistor measurements have been shown to detect individual charging events, but their spatial resolution is insufficient to resolve individual quasiparticles \cite{Zhitenev2000Imaginglocalizedelectronic,Ilani2004microscopicnaturelocalization,Martin2004LocalizationFractionallyCharged,Venkatachalam2011Localcharge5}. Direct real-space probes of individual quantum Hall quasiparticles are essential for understanding their structure and interactions, how they are bound and stabilized, and how they might ultimately be manipulated for topological quantum computing.

Scanning tunneling microscopy and spectroscopy (STM/STS) provide a powerful platform for probing electronic states with atomic spatial resolution. Recent STM studies of graphene in high magnetic fields have enabled, for the first time, microscopic examination of quantum Hall states \cite{Miller2010Realspacemappingmagnetically, Song2010Highresolutiontunnellingspectroscopy,Mao2016Realizationtunableartificial, Chiu2025Highspatialresolution, feldman2016observation, randeria2018ferroelectric}. Significant progress has been made in resolving broken valley symmetries \cite{liuVisualizingBrokenSymmetry2022,Coissard2022Imagingtunablequantum}, revealing spectroscopic characters of fractional quantum Hall states \cite{farahiBrokenSymmetriesExcitation2023a,Hu2025Highresolutiontunnellingspectroscopy}, detecting spatially ordered states \cite{tsuiDirectObservationMagneticfieldinduced2024}, and imaging edge-state structures \cite{Kim2021Edgechannelsbrokensymmetry, Yu2025Visualizinginteractiondrivenrestructuring, johnsen2023mapping}. However, microscopic signatures of individual quantum Hall quasiparticles—and their localization condition—remain unexplored.

Here, we employ scanning tunneling microscopy and spectroscopy (STM/STS) to directly visualize individual quantum Hall quasiparticles in graphene. By mapping LL energies in the $\nu = 2$ quantum Hall state, we reconstruct the real-space electrostatic potential landscape created by charged impurities. We show that strong near-surface impurity potentials lift Landau orbital degeneracy, enabling direct imaging of individual Landau orbitals. In the interaction-driven $\nu = -1$ quantum Hall ferromagnet state, we uncover distinct spectral signatures corresponding to zero to two trapped quasiparticles, controlled by defect depth and tunneling condition. Extending this approach to the $\nu = 5/3$ fractional quantum Hall state, we uncover spectroscopic signatures consistent with three bound anyonic quasiparticles. Our results establish charged defects as nanoscale traps for individual quantum Hall quasiparticles and demonstrate STM/STS as a platform for detecting their presence.

\phantomsection
\addcontentsline{toc}{subsection}{Imaging potential landscape and charged defects in graphene}

\section*{Imaging potential landscape and charged defects in graphene}

Our experiments are performed on monolayer graphene devices on hexagonal boron nitride (hBN) substrates with graphite back gates (inset of Fig. \ref{fig:1}a, details see \hyperref[sec:methods]{Methods}). The back-gate voltage, $V_\text{G}$, tunes the LL filling factor, $\nu = \frac{n h}{e B} $, where $n$, $h$, $e$, $B$ are the carrier density, Planck's constant, elementary charge and out-of-plane magnetic field. This work focuses on the spin- and valley-degenerate N = 0 Landau level (ZLL), spanning $\nu = -2$ to $\nu = 2$. For STS, we carefully prepare and select charge-neutral tips to minimize tip-induced gating effects (\hyperref[sec:methods]{Methods}). Gate-dependent spectra reveal a rich hierarchy of fractional quantum Hall states, attesting to the high quality of our samples (Supplementary Fig. S1). 

At the single-particle quantum Hall state of $\nu = 2$, the tunneling spectrum (Fig. \ref{fig:1}a) exhibits sharp spectroscopic peaks associated with the highly degenerate ZLL and N = 1 LL, separated by a cyclotron gap that spans the Fermi energy. The energy of these peaks is highly sensitive to the local electrostatic potential landscape in an incompressible quantum Hall state.
Fig.\ref{fig:1}c shows the spatial variation of the ZLL peak energy along a 500 nm line cut across the graphene surface at $\nu = -2$. By tracking the extracted peak positions of the ZLL, we map the energy shifts induced by local electrostatic potential variations over a 500$\times$500 nm$^2$ area in Fig. \ref{fig:1}b. These measurements reveal electrostatic variation on the scale of tens of meV.  Importantly, these features do not imply poor sample quality: topographic images acquired in the same region exhibit less than 1 nm of height variation with no observable surface contamination, while the weak topographic corrugation shows little correlation with the potential variations (Supplementary Fig. S2).

There are three clear circularly symmetric features in Fig. \ref{fig:1}b that originate from charged defects in graphene and the hBN substrate. These defects generate funnel-like potential profiles (e.g., Fig. \ref{fig:1}c) that follow the Coulomb potential of a $+e$ point charge. For the surface defects, topographic features are observed (Fig. \ref{fig:1}d), including a star-shaped pattern arising from intervalley scattering---an indication of the defect's atomic size\cite{Pereira2008Modelingdisordergraphene,Ugeda2010MissingAtomSource, Dutreix2019MeasuringBerryphase}. As shown later, its potential agrees well with that expected for a $+e$ charge defect on the graphene surface. For the subsurface defects, topographic signatures are barely or not observed (Fig. \ref{fig:1}e); however, their potential profiles are consistent with $+e$ or $-e$ charged defects buried several nanometers beneath the surface. We believe the random potential fluctuations in Fig. \ref{fig:1}b are a cumulative effect of Coulomb potentials from a distribution of deeper defects embedded within the hBN substrate \cite{zhang2009origin}. In the following discussion, we focus on phenomena induced by surface or shallow subsurface defects whose potentials can be individually resolved.

\phantomsection
\addcontentsline{toc}{subsection}{Orbital splitting and visualization of Landau orbital splitting at v = 2}

\section*{Orbital splitting and visualization of Landau orbitals at $\bm \nu$ = 2}

A high-resolution STS measurement along a line crossing a surface $+e$ defect reveals features beyond simple energy drifts (Fig. \ref{fig:2}a). As the tip approaches the defect center at $r = 0$, the ZLL peak shifts to lower energy following the Coulomb law but eventually splits into multiple branches. This behavior arises from lifting of the Landau orbital degeneracy by the defect potential. In the presence of a point-charge Coulomb potential, the symmetric-gauge wavefunctions, $\psi_m(r,\theta) \propto (\frac{r}{l_B})^m e^{im\theta} e^{-\frac{r^2}{4l_B^2}}$, remain eigenstates of the Hamiltonian, where $r$ is the distance from the defect, $\theta$ is the azimuthal angle, $m = 0, 1, 2...$ is the angular momentum quantum number, $l_B = \sqrt{\frac{h}{2\pi eB}}$ is the magnetic length. The Coulomb potential causes energy shifts that depend on $m$. The $m = 0$ orbital, being closest to the defect center, experiences the strongest potential and therefore exhibits the largest energy shift. Higher $m$ orbitals progressively trace the radial profile of the Coulomb potential of a $+e$ charged defect on the surface (green dashed curve in Fig. \ref{fig:2}a).

This effect, known as orbital splitting, had previously been only marginally resolved but manifests here with extraordinary clarity \cite{Luican-Mayer2014ScreeningChargedImpurities,Mao2016Realizationtunableartificial}. Each N = 0 orbital with $m > 0$ display a fine splitting into two peaks, arising from Zeeman-induced spin splitting while preserving valley degeneracy. Additionally, the $m=0$ orbital splits into four distinct levels. The upper two levels correspond to opposite spins within one valley, whereas the lower two arise from the opposite valley. The valley character is directly visible in the atomic-scale spatial patterns of the wavefunctions: the lower two peaks exhibit a lattice modulation shifted by a carbon bond relative to the upper two peaks (Fig. \ref{fig:2}a and Supplementary Fig. S3). The observation that valley splitting occurs only for $m=0$ points to valley symmetry breaking induced by atomic-scale interactions at the defect center, consistent with its absence for subsurface defects (Supplementary Fig. S4). The cusp-like feature near $r=0$ arises from tip screening: as the metallic tip approaches the defect, it partially screens the Coulomb potential, reducing the resulting energy shift. This effect is captured by a simple electrostatic model (Supplementary Fig. S5, and Supplementary Information Sec. 2).

Orbital splitting provides a direct way to visualize Landau orbitals. By integrating  d$I$/d$V$ over the energy range associated with the individual $m=0,1, 2$ orbitals, we obtain two-dimensional images of their probability densities (Fig. \ref{fig:2}b). The resulting patterns exhibit the expected sequence of rings with increasing spatial extent, the profiles of which are in good agreement with the theoretical expectation of $|\psi_m|^2$ (Fig. \ref{fig:2}c).

\phantomsection
\addcontentsline{toc}{subsection}{Spectroscopic signatures of quasiparticle binding at v = -1}

\section*{Spectroscopic signatures of quasiparticle binding at $\bm \nu$ = -1}

As the system moves from the cyclotron-gap state at $\nu = 2$ to the quantum Hall ferromagnetic state at $\nu = -1$, the same surface defect no longer simply alters the energies of uniformly filled Landau orbitals, but instead pins quantum Hall quasiparticles. At $\nu = -1$, each Landau orbital is singly occupied by a spontaneously spin- and valley-polarized state, denoted here as the $K\!\!\uparrow$ state. In the tunneling spectrum (Fig. \ref{fig:3}a), the occupied $K\!\!\uparrow$ state appears as a peak in the hole excitation channel ($V_\text{B} < 0$), whereas tunneling into the empty spin and valley states of the ZLL gives rise to two peaks at electron excitation ($V_\text{B}>0$) above the quantum Hall ferromagnetic gap.  The two electron excitation peaks correspond to different valleys and are split by valley-dependent electron-electron interactions associated with the shape of the Wannier orbitals \cite{PhysRevB.76.195415,PhysRevB.85.155439}. 

Focusing first on the hole excitation, we observe that the first indication of quasiparticle pinning arises from the spatial evolution of the filled ZLL peak (Fig. \ref{fig:3}a, left, $V_\text{B}<0$). Instead of shifting to lower energy following the Coulomb potential when the tip approaches the defect center, as observed in Fig. \ref{fig:2}a, the ZLL peak at $\sim$ -45 mV remains nearly flat and even shifts slightly upward. This behavior suggests that the Coulomb potential of a $+e$ defect is being screened by an oppositely charged quantum Hall quasiparticle. Assuming a $-e$ charged electron bound to the defect with the charge distribution of the $m=0$ state, the resulting screened potential reproduces the observed trend (green dashed curve in Fig. \ref{fig:3}a). In the absence of disorder, the elementary charged excitations of the $\nu = -1$ state are skyrmions --- topological spin textures carrying charge $\pm e$ distributed over a large area to reduce exchange energy cost \cite{Sondhi1993Skyrmionscrossoverinteger, MacDonald1996SkyrmionsSigmaModels, Yang2006Collectivemodesskyrmion}. Here, however, the strong defect potential favors a tightly bound state, driving the electron toward a minimal-size $m=0$ wavefunction. This electron is presumed to occupy the flipped-spin $K\!\!\downarrow$ state.

The right half of Fig. \ref{fig:3}a shows the calculated hole tunneling spectrum with a localized $K\!\!\downarrow$, $m=0$ electron. For this calculation, we find the excited states containing one fewer electron and calculate the overlap between the initial and final states (see details in Supplementary Materials). In the initial state, the $m = 0$ orbital is doubly occupied, containing both a $K\!\!\uparrow$ and $K\!\!\downarrow$ electron.  All other orbitals are singly occupied, each containing a single $K\!\!\uparrow$ electron. A tunneling event that removes the localized $K\!\!\downarrow$ electron produces a spectrally sharp excitation corresponding to the peak near -45 meV observed in the $m = 0$ orbital in both experiment and theory. We interpret the remaining spectral features as coming from states accessed when a $K\!\!\uparrow$ electron is removed at the tip location.  When the tip is far from the impurity, the resulting hole is largely localized on a large $m$ orbital, and its energy follows the combined Coulomb potential from the impurity and the bound electron (dashed green line in Fig. \ref{fig:3}a).  Removing a $K\!\!\uparrow$ electron in the $m = 0$ state creates a uniform density state with a flipped spin at the center (Supplementary Fig. S7k\&m).  This is a superposition of many different plane-wave spin-wave excitations (magnons), each with different energies. Consequently, the theoretical model predicts a broad spectral feature extending from roughly -50 meV to $-100$ meV.  The experiment instead shows a set of discrete peaks, suggesting that certain standing waves are stabilized.

The tunneling spectrum obeys a sum rule, namely the spectral weight, defined as the integrated d$I$/d$V$ over a spectral feature, is proportional to
electron/hole density in the corresponding mode $\rho(r)$,
\begin{align*}
\int_{V_1}^{V_2}\mathrm{d}I/\mathrm{d}V(r,V_\text{B})\,\mathrm{d}V_\text{B} \propto \rho(r)
\end{align*}
The curve represented by red markers in Fig. \ref{fig:3}b shows the spectral weight extracted over an energy window (details are shown in \hyperref[sec:methods]{Methods} and Supplementary Fig. S12) encompassing the removal spectrum of the $K\!\!\downarrow$, $m=0$ electron and the $K\!\!\uparrow$, $m>0$ electrons, excluding broad spectral features attributed to the removal of $K\!\!\uparrow$, $m = 0$ electron. The resulting density is uniform and in good agreement with the theory prediction (solid red curve). 

Switching to a subsurface defect (Fig. \ref{fig:3}c), we find that, under the same experimental conditions (hole tunneling at $\nu=-1$), no electron binding occurs. Instead, the orbital splitting pattern (Fig. \ref{fig:3}c) follows the bare Coulomb potential of a $+e$ point charge buried $\sim$ 5 nm beneath graphene (green dashed line in Fig. \ref{fig:3}c, Supplementary Fig. S14c\&d) and closely matches the STS observed at $\nu = 2$ for the same defect (Supplementary Fig. S4).  Extracting the spectral weight over $V_\text{B}$ reveals a uniform density of filled states (red markers in Fig. \ref{fig:3}d).  These observations demonstrate that the weaker Coulomb potential of the subsurface defect is insufficient to overcome the quantum Hall ferromagnetic gap and nucleate an electron. The stark dichotomy between the STS of surface and subsurface defects—reflecting distinct regimes of quasiparticle pinning—establishes a “quasiparticle microscope” capable of detecting the presence of localized quantum Hall quasiparticles. 

For each of these defects, when tunneling electrons ($V_\text{B} > 0$), we find STS is consistent with one extra defect-pinned electron, compared to the hole-excitation scenario discussed above. For the surface defect (Fig. \ref{fig:3}a), the energy of electron-tunneling spectral peak rises considerably upon approaching the impurity, consistent with what one would expect if two electrons were bound to the impurity.  There is a good agreement between the observed spectrum and a model with two bound electrons  (Fig. \ref{fig:3}a, right), and stark disagreement with what one would expect if only a single electron was bound (Supplementary Fig. S7b). The spectral weight of the electron excitations, proportional to the density of holes, is strongly suppressed near the defect center (blue markers in Fig. \ref{fig:3}b), in agreement with the theoretical prediction for two pinned electrons (blue line). Second electron pinning is enabled by the imperfect screening of the defect potential by only one single electron, which remains attractive near the defect center (green dashed line in Fig. \ref{fig:3}a).  Similar physics allows two electrons to be bound to a single proton, forming a H$^-$ ion. 
 
For the subsurface defect, the electron excitation STS (Fig. \ref{fig:3}c) likewise deviates from what one would expect from a bare Coulomb potential and instead resembles the spectrum of a single-electron bound state (Fig. \ref{fig:3}c, right; Supplementary Fig. S7g\&h). We note residual discrepancies between experiment and theory in the electron excitation, which may be attributed to the nature of the additional pinned electron. The theory assumes a minimal $m = 0$ orbital wavefunction, whereas a small skyrmion may instead be stabilized in the screened and subsurface defect potentials. The inferred quasiparticle binding configurations for the two defects under different tunneling conditions are summarized in Fig. \ref{fig:3}f. 
As we elaborate on below, the act of tunneling electrons/holes into the sample changes the occupancy of the defect bound state.
The STS at other quantum Hall ferromagnetic states, including $\nu = 0$ and $1$, can be understood similarly with different numbers of pinned quasiparticles (Supplementary Fig. S4).

We present a unified picture that explains the conditions for quasiparticle pinning and why the same impurity exhibits markedly different trapping behavior for electron and hole excitations. To illustrate, we show a schematic energy-level diagram in Fig. \ref{fig:3}e. The Coulomb potential of a $+e$ defect creates a localized $K\!\!\downarrow$ state within the gap with binding energy $E_{m = 0}$. When the Fermi energy lies above this localized state, the corresponding quasiparticle state is occupied, and vice versa. In transport experiments, the Fermi energy—and consequently the quasiparticle occupation—can be tuned by gate voltage. Surprisingly, in our experiment, varying the gate voltage within the incompressible region does not alter the quasiparticle occupation (Supplementary Fig. S8 and Supplementary Videos). To understand this behavior, one must consider the effect of the tunneling current from the STM tip, and the abundance of localized in-gap states.

Incompressible quantum Hall states are insulating in the bulk. During STM/STS measurements, tunneling electrons or holes accumulate in localized in-gap states (light blue or orange in Fig. \ref{fig:3}e), causing the local chemical potential to shift until it reaches the mobility edge of the extended states (dark blue and red). As a result, the tunneling process effectively pins the local chemical potential to the top of the gap at $\mu_{\mathrm{bulk}}^e$ for electron tunneling and at the bottom of the gap at $\mu_{\mathrm{bulk}}^h$ for hole tunneling (Fig. \ref{fig:3}e), rendering gate-controlled chemical potential only effective at the sample edge and irrelevant in areas around the tunneling site. Within this picture, in hole excitation, an electron binds to the $+e$ defect only when the binding energy $E_{m = 0}$ exceeds the quantum Hall gap $\Delta_\text{c}$ (energy cost for creating an electron), whereas electron tunneling always results in at least one bound electron state to an attractive defect potential. This tunneling-induced pinning of the local chemical potential also explains the apparently enlarged quantum Hall gaps observed in incompressible states (Supplementary Fig. S9).

\phantomsection
\addcontentsline{toc}{subsection}{Signatures of anyon bound states at v = 5/3}

\section*{Signatures of anyon bound states at $\bm \nu$ = 5/3}

Now we turn our attention to fractional quantum Hall states at $\nu = \pm 5/3$, corresponding 
to $1/3$ filling of electrons or holes within a single spin- and valley-polarized branch of the ZLL, which are the simplest fractional quantum Hall states in monolayer graphene. Here, depending on the strength of the Coulomb potential, an impurity is expected to bind zero, one, two, three, or more anyons, each carrying charge $\pm e/3$, in contrast to integer quantum Hall states, where the quasiparticle charge changes in units of $\pm e$. Our theoretical density matrix renormalization group (DMRG) calculations reveal distinct spectroscopic structures associated with different numbers of bound anyons, reflected both in the orbital-resolved spectral features and in the screened potential profile away from the defect (the three anyon case is shown in Fig. \ref{fig:4}b, the rest are shown in Supplementary Fig. S10).

Experimentally, we focus on the electron excitation spectrum of $\nu = 5/3$, which probes the $1/3$-filled spin- and valley-polarized hole state (Fig. \ref{fig:4}a), avoiding complications from additional spin and valley degrees of freedom. The measured STS (Fig. \ref{fig:4}c) near a $-e$ charged defect exhibits remarkable agreement with the theoretical prediction for a bound state containing three $+e/3$ anyon quasi-holes (Fig. \ref{fig:4}b). In particular, both the screened Coulomb potential away from the defect and the energy sequence of $m = 0, 1, 2$ spectral peaks are accurately reproduced, whereas simulations for zero, one, and two bound anyons display clear discrepancies (Supplementary Fig. S10). Measurements at $\nu = -5/3$ near a $+e$ charged defect reveal nearly identical spectral features in the hole-excitation channel (Supplementary Fig. S11), mirroring the data in Fig. \ref{fig:4}c. This particle-hole symmetry further supports the robustness of our spectroscopic observation in fragile fractional quantum Hall states.

The spectral weight of the electron excitation at $\nu=5/3$ reveals the real-space density profile of the three anyons quasi-hole bound state (Fig. \ref{fig:4}f, g). Expressed in units of hole filling per Landau orbital, the spectral weight (Fig. \ref{fig:4}g) exhibits a pronounced peak near $r=0$, approaching unity, indicating nearly complete depletion of the local $K'\!\!\downarrow$ orbitals. The spatial extent of this peak exceeds that of the $m=0$ orbital probability density. In addition, oscillations in the hole density are observed several $l_B$ away from the defect.
 These oscillations are also evident in the two-dimensional map of the excess charge density $\Delta\rho$ shown in Fig. \ref{fig:4}f, which provides a real-space image of the charge distribution associated with the three-anyon bound state. For a weak defect potential, such oscillatory structures can be related to the magneto-roton spectrum of collective excitations \cite{Girvin1986Magnetorotontheorycollectivea}. These nontrivial spatial charge features provide further evidence that the bound state is not a simple $m=0$ hole despite carrying a total charge of $+e$. Rather, it can be viewed as an example of an anyon molecule \cite{gattu2025molecular, wang2026anyon}.

Lastly, we examine the properties of a $+e$ defect in the same $\nu = 5/3$ fractional quantum Hall state (Fig. \ref{fig:4}e). Upon approaching the defect center, all electron-tunneling spectral features disappear, and the spectral weight drops to zero (Fig. \ref{fig:4}f). This behavior reflects the presence of bound quasi-electrons at the positively charged defect, which locally produce a fully filled ZLL, completely suppressing electron tunneling. In this case, information about the number of trapped anyon quasi-electrons is encoded primarily in the spatial extent of the spectral-weight suppression, rather than in discrete spectroscopic features as observed for bound anyon quasi-holes (Supplementary Fig. S10). Consequently, extracting the number of bound anyons is less reliable. Nevertheless, our best estimate suggests that three anyon quasi-electrons, carrying a total charge of $-e$, are localized at the defect (Fig. \ref{fig:4}d and Supplementary Fig. S10).

\phantomsection
\addcontentsline{toc}{subsection}{Conclusion and outlook}

\section*{Conclusion and outlook}

Our work establishes STS as a powerful platform for detecting and visualizing localized quasiparticles in both integer and fractional quantum Hall states. In the future, examining defects buried deeper from the graphene surface may enable the observation of one- and two-anyon bound states, while similar techniques could be extended to probe quasiparticles in non-Abelian fractional quantum Hall states and fractional Chern insulators. The STM platform also opens the possibility of combining anyon visualization with active manipulation through tip-induced electrostatic potentials, raising the prospect of demonstrating anyon braiding. Beyond quasiparticle detection, our experiments provide insight into the presence and effects of charged defects in high-quality van der Waals heterostructures and reveal how tunneling processes affect spectroscopic measurements in insulating states.
\clearpage

\phantomsection
\addcontentsline{toc}{section}{Figures}
\phantomsection
\addcontentsline{toc}{subsection}{Fig. 1 \textbar{} Spectroscopic imaging of the zeroth Landau level at v = 2, revealing potential landscape and charged defects.}

\begin{figure*}
\refstepcounter{figure}
\label{fig:1}
\noindent\makebox[\textwidth][c]{%
  \includegraphics[width=\linewidth]{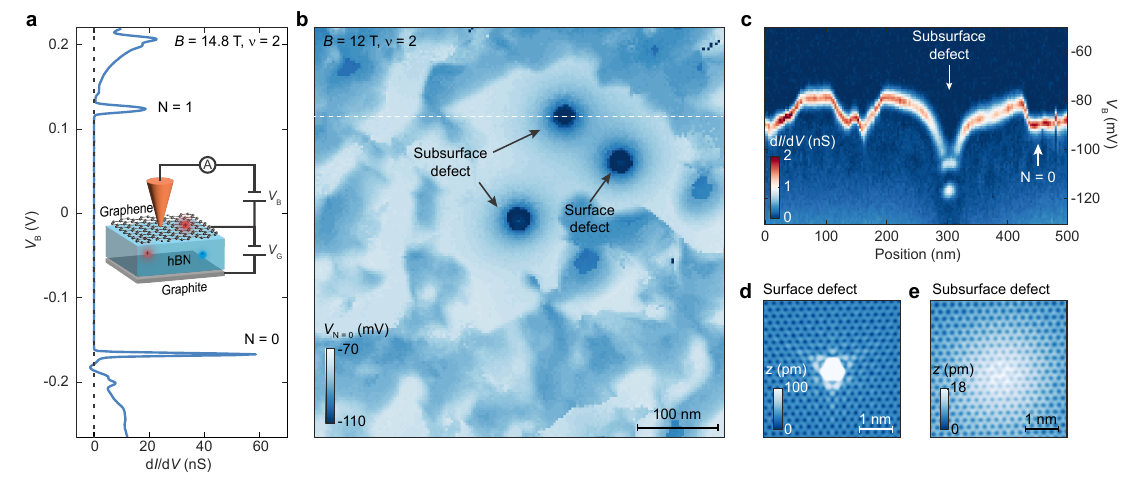}
}
\end{figure*}
\vspace{0pt}

\noindent \textbf{Fig. \ref{fig:1} \textbar{} Spectroscopic imaging of the zeroth Landau level at $\nu = 2$, revealing potential landscape and charged defects.}
\textbf{a}, Representative tunneling spectrum in the $\nu = 2$ (gate voltage $V_\text{G}=1.416$ V) quantum Hall state at $B = 14.8$ T shows spectral peaks corresponding to the N = 0, 1 Landau levels (LLs) and the gap in between. Inset: schematic of the experimental setup, where $V_\text{B}$ and $V_\text{G}$ are the bias and gate voltage, respectively.
\textbf{b}, Real-space map of the zeroth Landau level (ZLL) peak bias voltage extracted from STS measurements over 500$\times$500 nm at $B = 12$ T, $\nu = 2$, and $V_\text{G}=1.107$ V. Peak shifts reveal local electrostatic potential variations.
\textbf{c}, Position-dependent d$I$/d$V$ along a line (white dashed line in \textbf{b}) that crosses a shallow subsurface $+e$ defect. 
\textbf{d}, \textbf{e}, Representative atomically resolved topography images of a surface (\textbf{d}, setpoint voltage $V_\text{B} = -200$ mV, setpoint tunneling current $I_{\mathrm{t}} = 200$ pA) and a subsurface (\textbf{e}, $V_\text{B} = -500$ mV, $I_\text{t} = 200$ pA) defect at $\nu=2$. Note the difference in $z$ scale.

\clearpage

\phantomsection
\addcontentsline{toc}{subsection}{Fig. 2 \textbar{} Orbital splitting at v = 2 and visualization of Landau orbits.}

\refstepcounter{figure}
\label{fig:2}
\noindent\makebox[\textwidth][c]{%
  \includegraphics[width=0.65\textwidth]{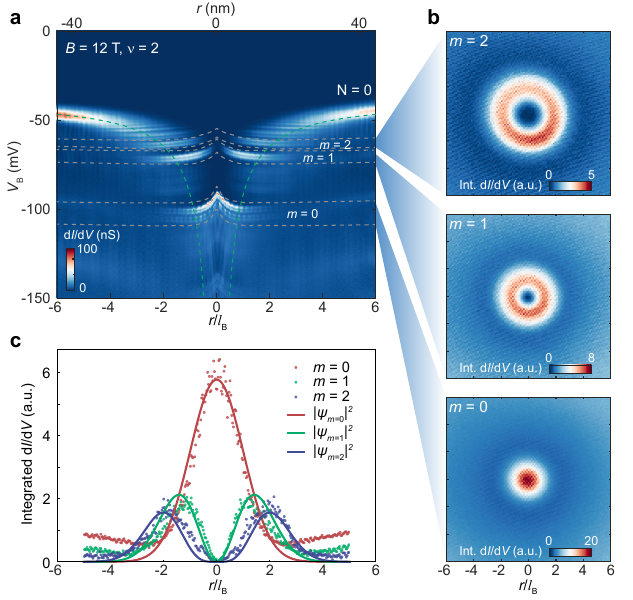}%
}
\vspace{0pt}

\noindent\textbf{Fig. \ref{fig:2} \textbar{} Orbital splitting at $\nu = 2$ and visualization of Landau orbits.}
\textbf{a}, High-resolution STS across a surface $+e$ charged defect at $B = 12$ T, $\nu = 2$ ($V_\text{G}=1.065$ V), showing the energy splitting of Landau orbitals according to their angular momentum quantum number $m$. The tip height is set by $V_\text{B}=-200$ mV, $I_\text{t}=2$ nA. 
The energies follow the Coulomb potential profile of a $+e$ surface defect (green dashed curves). The cusp at $r = 0$ is caused by tip screening (see Supplementary Fig. S5).
\textbf{b}, Maps of integrated d$I$/d$V$ over energy windows that cover $m = 0, 1, 2$ Landau orbitals, reflecting the probability density of these states. The dataset is acquired at $\nu = 2$ at $B = 14.8$ T (see Supplementary Fig. S4). We use a position-dependent energy window that tracks the corresponding spectral peak, as illustrated by the gray dashed curves in \textbf{a}. 
\textbf{c}, Comparison of the probability density extracted from integrated experimental d$I$/d$V$ (dots) with theoretical probability density of Landau orbitals (solid curves) along a line cut across the defect.

\clearpage

\phantomsection
\addcontentsline{toc}{subsection}{Fig. 3 \textbar{} Spectroscopic signature of bound quasiparticles in the v = -1 quantum Hall ferromagnetic state.}

\refstepcounter{figure}
\label{fig:3}
\noindent\makebox[\textwidth][c]{%
  \includegraphics[width=\textwidth]{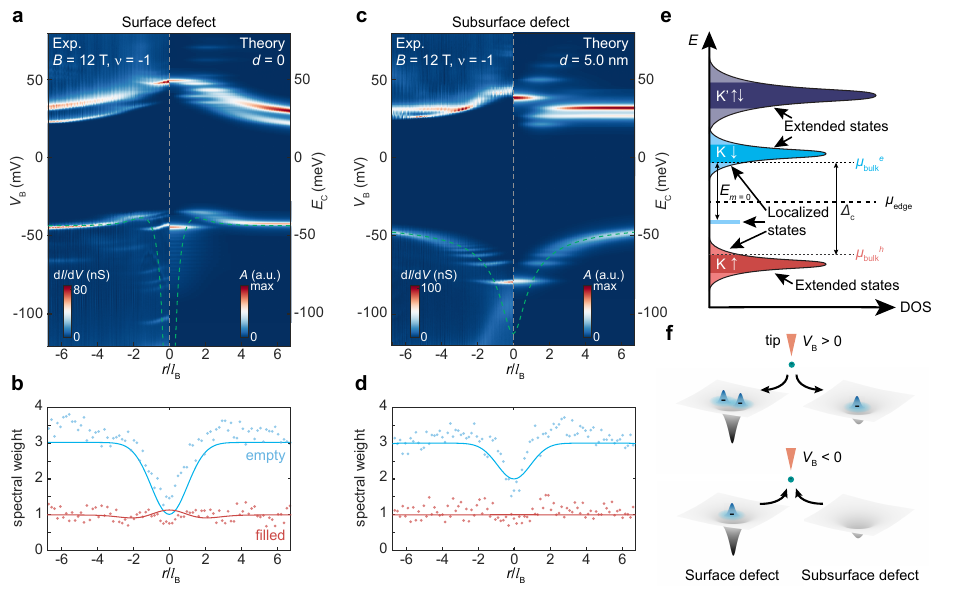}%
}
\vspace{0pt}

\noindent\textbf{Fig. \ref{fig:3} \textbar{} Spectroscopic signature of bound quasiparticles in the $\nu = -1$ quantum Hall ferromagnetic state.}
\textbf{a}, \textbf{c}, Experimental STS (left) across a surface and a subsurface $+e$ surface defect at $B = 12$ T and $\nu = -1$ compared with corresponding theoretical calculated spectral function (right). For the surface defect, the theory model assumes that the initial state contains one bound $m = 0$ electron for hole tunneling ($V_\text{B}<0$) and two bound $m = 0$ electrons for electron tunneling ($V_\text{B}>0$). For the subsurface defect, the model contains zero and one electron for hole and electron tunneling, respectively. The green dashed curves represent the screened potential composed of a surface $+e$ defect and a bound $m=0$ electron in \textbf{a}, and the bare Coulomb potential of a subsurface $+e$ charge at depth $d=5$ nm in \textbf{c}. The tip heights are set by $V_\text{B}=-200$ mV, $I_\text{t}=2$ nA for both spectra.
\textbf{b}, \textbf{d}, Spectral weight for electron- (blue symbols) and hole-excitation (red symbols) corresponding to the surface and subsurface defects shown in \textbf{b} and \textbf{d} respectively. Dots represent the experimental data, while the solid curves are obtained from theoretical calculations. The spectral weight is extracted by integrating d$I$/d$V$ over a finite energy window and normalized such that the sum of filled and empty states far from the defect equals four states per Landau orbital (see \hyperref[sec:methods]{Methods} and Supplementary Fig. S12 for analysis details). For the hole-excitation across the surface defect (red symbols in \textbf{b}), a narrow integration window below the quantum Hall ferromagnetic gap is chosen to isolate the removal spectrum of the trapped electron while excluding the less well-understood features at $V_\text{B} < -55$ mV. All other spectral weight plots incorporate all LL features with either positive or negative energy. 
\textbf{e}, Illustration of chemical potential pinning induced by tunneling. The density of states (DOS) of the ZLL is schematically shown as a function of energy. A quantum Hall ferromagnetic gap, $\Delta_\text{c}$, separates the filled $K\!\!\uparrow$ states from the empty $K\!\!\downarrow$ states. The energy separation between the empty $K$ and $K'$ valley states reflects the valley splitting arising from lattice-scale interactions. Owing to potential variations, Landau orbitals near the edge of LL (represented by the light blue and orange regions) are localized, trapping any incoming electrons or holes. A strong $+e$ charged defect potential can generate an isolated in-gap bound state (the light blue line) with binding energy $E_{m=0}$. Normally, the occupation of this defect-bound state is determined by the gate voltage, which sets $\mu_\text{edge}$. However, for electron and hole tunneling spectroscopy, the injected electrons or holes drive the local chemical potential toward the mobility edges, marked by $\mu_\text{bulk}^{e}$ and $\mu_\text{bulk}^{h}$ for electron and hole tunneling, respectively, regardless of the value of $\mu_\text{edge}$, provided it lies within the quantum Hall ferromagnetic gap.
\textbf{f}, Schematic illustrating the quasiparticle binding scenarios for surface (\textbf{a}) and subsurface defects (\textbf{c}) under electron (top row) and hole (bottom row) tunneling from the tip. Each blue cloud represents a trapped electron.


\clearpage

\phantomsection
\addcontentsline{toc}{subsection}{Fig. 4 \textbar{} Evidence for multiple bound anyonic quasiparticles in the fractional quantum Hall state at v = 5/3.}

\refstepcounter{figure}
\label{fig:4}
\noindent\makebox[\textwidth][c]{%
  \includegraphics[width=\textwidth]{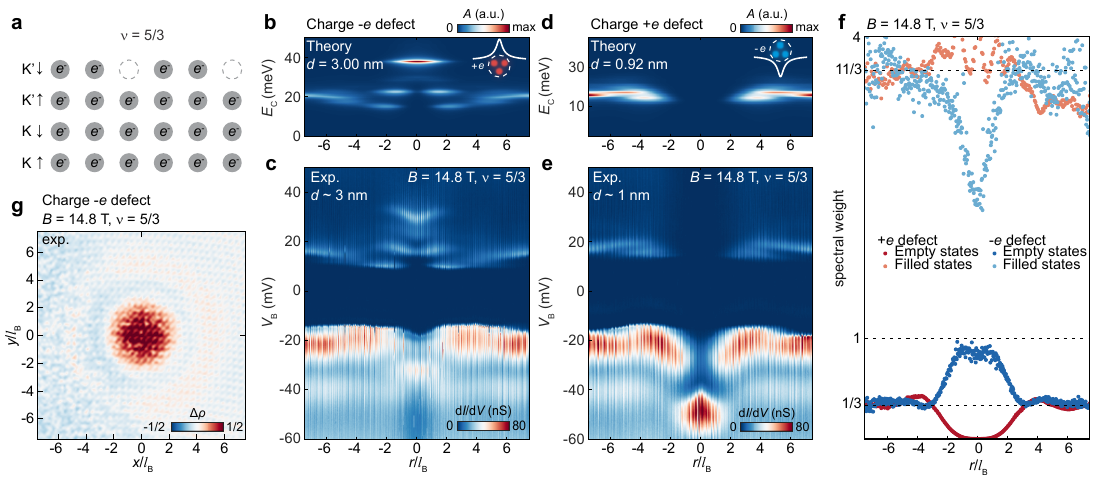}
  }
\vspace{0pt}

\noindent\textbf{Fig. \ref{fig:4} \textbar{} Evidence for three anyon bound states in the fractional quantum Hall state at $\nu = 5/3$.}
\textbf{a}, Schematic of the $\nu=5/3$ fractional quantum Hall state, corresponding to a hole filling of $\nu^* =1/3$ in the $K'\!\!\downarrow$ flavor of the ZLL. Solid and dashed circles indicate filled and empty states. 
\textbf{b}, \textbf{d}, DMRG simulations of the spectral function for the $\nu^*=1/3$ fractional quantum Hall state with three bound anyon quasi-holes (\textbf{b}) and quasi-electrons (\textbf{d}) around $-e$ and $+e$ defects with depths of $3$ nm and $0.92$ nm, respectively. Insets illustrate three bound anyons, carrying a total charge of $\pm e$, confined by the Coulomb potentials of the $\mp e$ defects (white curves).
\textbf{c}, \textbf{e}, Experimental STS measured across a negatively charged $-e$ subsurface defect with depth $d\approx3$ nm (\textbf{c}) and a positively charged $+e$ subsurface defect with depth $d\approx1$ nm (\textbf{e}) at $B=14.8$ T and $\nu=5/3$. Measurement conditions are described in the caption of Supplementary Fig. S12. 
\textbf{f}, Spectral weight extracted from the experimental STS data in \textbf{c} and \textbf{e}. Dark (light) markers denote empty-state (filled-state) spectral weights, with blue and red corresponding to the $-e$ and $+e$ defects, respectively. The spectral weight is normalized to represent the local electron or hole filling per Landau orbital (see Supplementary Fig. S12 and \hyperref[sec:methods]{Methods} for analysis details). The local enhancement or suppression of the spectral weight near the defect center reflects the presence of trapped quasiparticles.
\textbf{g}, Real-space image of the empty-state spectral-weight variation $\Delta\rho$ at $B=14.8$ T and $\nu=5/3$ around the $3$ nm deep $-e$ defect. Here, $\Delta\rho$ is defined as the spectral weight, normalized to represent the hole filling per Landau orbital, with the uniform background value of $1/3$ subtracted. It therefore represents the excess hole density associated with quasiparticle binding relative to the $\nu=1/3$ Laughlin-state background.


\clearpage





\section*{Methods}
\label{sec:methods}
\phantomsection
\addcontentsline{toc}{section}{Methods}


\section*{Sample fabrication}
\label{sec:sample_fabrication}
\phantomsection
\addcontentsline{toc}{subsection}{Sample fabrication}

Graphene devices in this experiment are fabricated with a method similar to previous reports \cite{liuVisualizingBrokenSymmetry2022}. Monolayer graphene and hBN flakes were mechanically exfoliated on SiO$_2$ (285 nm)-Si wafers. The graphene heterostructures were made using polyvinyl alcohol-coated transparent tape with a polydimethylsiloxane block on a glass slide as a support. Graphene, contact graphite, hBN, and bottom gate graphite layers were sequentially picked up and finally released onto a SiO$_2$-Si substrate with pre-patterned Au/Cr contact electrodes. The finished devices were washed with HPLC water, acetone, isopropyl alcohol, and $n$-methyl-2-pyrrolidone. The devices were then annealed first in a forming gas (hydrogen/nitrogen) at around 320 \textdegree C and then in an ultrahigh vacuum chamber at around 350 \textdegree C for more than 10 hours before performing STM experiments.

\section*{STM and STS measurement}
\label{sec:stm_methods}
\phantomsection
\addcontentsline{toc}{subsection}{STM and STS measurement}

All experimental results in this work were acquired using a Unisoku 1300 STM system at a base temperature of $T \sim 400$ mK under high out-of-plane magnetic fields ranging from $B = 12$ to $14.8$ T. Spectroscopy data were measured by the standard lock-in technique with $f_\text{ref}=932$ Hz at modulation voltage $V_\text{rms}=0.2\sim 1$ mV. To reduce measurement noise at small bias voltages, we employed a 1:10 bias divider, which attenuates the output bias by a factor of 10 before it is applied to the sample. All bias voltages reported in this work are given after the divider. We located the micrometer-size sample using a capacitance-guiding method \cite{liSelfnavigationScanningTunneling2011}. We monitor the tip-sample capacitive current with a lock-in excitation applied between the tip and sample ($V_\mathrm{rms}=2$ V, $f_\mathrm{ref}=8.311$ kHz), while scanning the tip above the sample at a height of about 5 - 20 $\mu$m. The graphite gate and silicon substrate are kept grounded in this process.

Before measurements, we calibrate the tip on an Ar-ion sputtered and annealed Cu(111) single crystal with an atomically flat surface. We prepare the STM tip by repeatedly poking it into the Cu substrate until a small poke mark is produced and a clean spectrum showing the Cu(111) surface state is obtained. We then measure gate-dependent spectrum in the graphene quantum Hall system and only proceed when Haldane sash features are observed \cite{farahiBrokenSymmetriesExcitation2023a}, indicating a minimally invasive and effectively charge-neutral tip. Such a tip condition is essential for reliably probing fragile symmetry-breaking and fractional quantum Hall states.

Supplementary Fig. S1 shows an optical image of the two devices used for this study and examples of their gate-dependent point spectra away from strong defect potentials. 
To obtain a high-quality gate-dependent point spectrum, we maintain a constant tip height throughout the measurement using a procedure similar to that described previously  \cite{liuVisualizingBrokenSymmetry2022}. This is achieved by first acquiring a reference tunneling current curve as a function of gate voltage $I_\text{t}$ ($V_\text{G}$) (Supplementary Fig. S13) while the tip feedback controller is off (constant height). This reference is then played back as the current setpoint when we obtain the gate-dependent spectrum that takes several hours to acquire. We define this procedure as the constant-height-setpoint approach.


For spectral-weight measurements across charged defects, additional care is needed.  It is crucial to maintain a constant tip height above the graphene surface as the tip scans through the defect. This is challenging because charged defects induce spatial variations in the local spectrum, particularly in incompressible states. If a fixed bias voltage and current setpoint are used to determine the tip height, the effective tip-sample distance will vary as the tip approaches the defect. As a result, a fixed setpoint parameter within the incompressible state does not guarantee a constant tip height during spatial scans across charged defects, which would compromise the accuracy of spectral-weight analysis.

To minimize this effect, we employ the following procedure. First, the tip height is set in a compressible state, where the defect potential is strongly screened. A large setpoint bias voltage encompassing multiple LLs is used to further reduce the influence of the defect on the tunneling current. The feedback loop is then turned off to maintain a constant tip height, and the gate voltage is rapidly ramped to the target value in the incompressible state before STS measurements are performed. The setpoint gate voltage, bias voltage, and tunneling current used for each measurement are specified in the corresponding figure captions.



\section*{Estimation of the subsurface defect depth}
\label{sec:defect_split}
\phantomsection
\addcontentsline{toc}{subsection}{Estimation of the subsurface defect depth}

We estimate the depth of subsurface $+e$ defects by fitting the experimental energy splitting between different angular momentum orbitals $m$, $m'$,  $\Delta E_{m, m'}$,  with a theoretical formula from the defect Coulomb potential:
\begin{align*}
      \Delta E_{m, m'} = \vert E_m - E_{m'}\vert\quad E_m = \langle \psi_m \vert \frac{-e^2}{4\pi \epsilon \epsilon_0 \sqrt{r^2 + d^2}} \vert \psi_m\rangle
\end{align*}
where $\psi_{m}$ is $m$ angular momentum LL eigenstate, $e$ is the elementary charge, $\epsilon$ and $\epsilon_0$ are the relative and vacuum dielectric constants, $r$ is the radius away from the defect center, and $d$ is the defect depth.

We fit the model to the experimental hole-tunneling spectrum ($V_\text{B}<0$) in the absence of quasiparticle binding, where the defect potential corresponds to the bare Coulomb potential. In the above formula, there are two parameters to be determined: $\epsilon$ and $d$. We first fit a surface defect (with known $d=0$) to obtain a best fit of $\epsilon\approx 3.88$. We then round the dielectric constant to $\epsilon=4$, and find $d$ as the only fitting parameter (results in Supplementary Fig. S14).

A special case is the subsurface $-e$ defect shown in Fig. \ref{fig:4}c. The defect exhibits a gate-switchable charge state, carrying a charge of $-e$ for $\nu>1.5$ and becoming charge neutral otherwise (Supplementary Fig. S13). Consequently, the $\nu=2$ state is the only regime in which the orbital splitting corresponding to a bare Coulomb potential can be used to extract the defect depth. The corresponding STS data are shown in Supplementary Fig. S13d, where the unscreened orbital splitting pattern of the N = 1 LL appears in electron excitation. Owing to the opposite defect charge and the chemical-potential pinning effect, this splitting pattern is inverted relative to that of a $+e$ defect.
We compare the inverted orbital splitting pattern (Supplementary Fig. S14j) with the hole-excitation STS of the N = -1 LL at $\nu=-2$ for other subsurface defects with known depths (Supplementary Fig. S14i and k), yielding an estimated depth of $d\approx3$ nm for this defect. We do not directly fit the data using the theoretical model for the N = 1 orbital splitting because tip screening and tip-gating effects are significantly stronger for the N = $\pm$1 LL (Supplementary Notes Sec. 2 and Fig. S5).

\section*{Acknowledgements}
\addcontentsline{toc}{section}{Acknowledgements}

We acknowledge helpful discussions with T. Wang, G. Murthy, C.L. Chiu, G. Farahi, A. Yazdani, and A. M. Turner. This work was supported by the National Science Foundation through CAREER Award No. DMR-2442363. E.J.M. and D.P. were supported by the National Science Foundation under NSF-2409403. J.D. acknowledges support from the New Frontier Grant, College of Arts and Sciences, Cornell University. K.W. and T.T. acknowledge support from the CREST (JPMJCR24A5), JST and World Premier International Research Center Initiative (WPI), MEXT, Japan. This work was performed in part at the Cornell NanoScale Facility, a member of the National Nanotechnology Coordinated Infrastructure (NNCI), which is supported by the National Science Foundation (Grant NNCI-2025233). 

\phantomsection
\addcontentsline{toc}{subsection}{Author contributions}
\paragraph*{Author contributions:}
X.L. conceived the project. Y.S. fabricated the device. J.D. and Y.S. performed the measurements, analyzed the data, and prepared the figures with input from X.L. E.J.M. and D.P. carried out the theoretical modeling and calculations. K.W. and T.T. provided hBN crystals. J.D., Y.S., E.J.M., and X.L. wrote the manuscript with input from all authors.

\phantomsection
\addcontentsline{toc}{subsection}{Competing interests}
\paragraph*{Competing interests:}
There are no competing interests to declare.

\clearpage
\phantomsection
\addcontentsline{toc}{section}{References}
\section*{References}
\bibliography{BIB_qhSTM}
\bibliographystyle{naturemag_noURL}


\clearpage

%

\clearpage

\setlength{\parindent}{0pt}
\setlength{\parskip}{0pt} 

\setcounter{figure}{0}
\setcounter{table}{0}
\setcounter{equation}{0}
\setcounter{subsection}{0}
\setcounter{subsubsection}{0}

\providecommand{\angstrom}{\mbox{\normalfont\AA}}
\providecommand{\citebox}[1]{\colorbox{blue}{\textcolor{white}{\textbf{#1}}}}
\providecommand\redsout{\bgroup\markoverwith{\textcolor{red}{\rule[0.5ex]{2pt}{0.4pt}}}\ULon}
\providecommand{\nutotal}{$\nu_{total}$}
\providecommand{\red}{\textcolor{red}}
\providecommand{\blue}{\textcolor{blue}}
\providecommand{\green}{\textcolor{green}}
\providecommand{\CF}{$_{1}^{2}$CF }
\providecommand{\CFs}{$_{1}^{2}$CFs }
\providecommand{\dlb}{$d/\ell_B$}
\providecommand{\up}{$|\uparrow >$}
\providecommand{\down}{$|\downarrow >$}
\providecommand{\ttCF}{$_{2}^{2}$CF }
\providecommand{\ttCFs}{$_{2}^{2}$CFs }
\providecommand{\tzCF}{$_{0}^{2}$CF }
\providecommand{\tzCFs}{$_{0}^{2}$CFs }
\providecommand{\LLL}{$\Lambda$-level }
\providecommand{\RCF}{$R_{xx}^{CF}$ }
\providecommand{\dragxy}{$R_{xy}^{drag}$ }
\providecommand{\dragxx}{$R_{xx}^{drag}$ }

\renewcommand{\thesubsubsection}{\arabic{subsubsection}}
\providecommand{\subsectionheading}[1]{%
    \stepcounter{subsection}%
    \subsection*{#1}%
}

\begin{center}
{\Large\bfseries Supplementary Information for}\\[0.5em]
{\Large\bfseries Real-space Imaging of Quantum Hall Quasiparticles}
\end{center}

\vspace{2em}

\phantomsection
\addcontentsline{toc}{subsection}{Table of Contents}
\noindent{\large\bfseries Table of Contents}\par





\noindent \textbf{Supplementary Notes}

\noindent \ref{subsubsec:theory_v-1} Simulation of spectra of the $\nu$ = -1 quantum Hall ferromagnetic state

\noindent \ref{subsubsec:tip-sim} Simulation of tip effects

\noindent \ref{subsubsec:tip-matrixelements} Calculating displaced Coulomb matrix elements

\noindent \ref{subsubsec:theory_v1/3} Simulation of spectra of the $\nu^*$ = 1/3 fractional quantum Hall state

\noindent \ref{subsubsec:coulomb_mat_defect} Numerically calculating Coulomb matrix elements with defect

\noindent \ref{subsubsec:coulomb_mat_electrons} Numerically calculating Coulomb matrix elements between electrons

\noindent \ref{subsubsec:trick_fractional} Calculating Coulomb matrix elements on a sphere

\noindent Fig. S\ref{fig:device_spectra} \textbar{} Device images and gate-dependent d\textit{I}/d\textit{V} spectra.

\noindent Fig. S\ref{fig:s2} \textbar{} Comparison between surface topography and simultaneously acquired potential landscape.

\noindent Fig. S\ref{fig:s3} \textbar{} Visualizing valley symmetry-breaking of \textit{m} = 0 orbital on a surface defect.

\noindent Fig. S\ref{fig:s4} \textbar{} STS line cuts across surface and subsurface defects at various integer fillings.

\noindent Fig. S\ref{fig:tip_sim} \textbar{} Electrostatic simulation of the tip effect on LL spectra.

\noindent Fig. S\ref{fig:s6} \textbar{} Spectral weight analysis across the surface +\textit{e} defect at $\nu$ = -1 with different integrating windows.

\noindent Fig. S\ref{fig:sim_v_-1} \textbar{} Simulated spectra for surface and subsurface +\textit{e} defect at $\nu$ = -1.

\noindent Fig. S\ref{fig:s8} \textbar{} Gate-dependence of STS line cuts spanning the $\nu$ = -1 quantum Hall ferromagnetic states for subsurface and surface defects at \textit{B} = 14.8 T in device A.

\noindent Fig. S\ref{fig:pinning} \textbar{} Tunneling-current-induced chemical potential pinning and its effect on the apparent quantum Hall gap in STS.

\noindent Fig. S\ref{fig:fractional_sim} \textbar{} DMRG simulation of $\mp$\textit{e} defect spectra at \textit{B} = 14.8 T, hole filling $\nu^*$ = 1/3 fractional quantum Hall state.

\noindent Fig. S\ref{fig:negative_inverted} \textbar{} Comparison between STS across a +\textit{e} charged defect at $\nu$ = -5/3 and a -\textit{e} charged defect at $\nu$ = 5/3, showing particle-hole symmetry.

\noindent Fig. S\ref{fig:s12} \textbar{} Spectral weight extraction for the spectral weight plots shown in Fig. 3 and 4.

\noindent Fig. S\ref{fig:gate_switch_defect} \textbar{} Characterization of the switchable subsurface -\textit{e} defect in Fig. 4.

\noindent Fig. S\ref{fig:defect_split} \textbar{} Estimation of subsurface defect depths.

\clearpage


\phantomsection
\addcontentsline{toc}{subsection}{Supplementary Notes}
\noindent{\large\bfseries Supplementary Notes}\par

\subsubsection{Simulation of spectra of the \texorpdfstring{$\bm \nu$}{v} = -1 quantum Hall ferromagnetic state}
\label{subsubsec:theory_v-1}

To calculate the spectral functions in the $\nu=-1$ state, we use ``exact diagonalization," where we exactly solve the many-body problem of a filled spin-polarized Landau level (LL) with a fixed number of extra particles and/or holes.  For this calculation, we work with an infinite plane, placing the charge $+e$ impurity at the origin and using the symmetric gauge for the magnetic field. The Hamiltonian describing the interactions between the electrons and the impurity, and the Coulomb interaction between electrons is then
\begin{align}
    H&= \sum_{\alpha, m} (U_m-\mu_\alpha) \hat \psi_{\alpha, m}^\dagger \hat \psi_{\alpha, m}
    +\frac{1}{2} \sum_{m_1 m_2 m_3 m_4}\sum_{\alpha, \beta} V_{m_1 m_2 m_3 m_4} \hat \psi_{\alpha, m_1}^\dagger\hat \psi_{\beta, m_2}^\dagger \hat \psi_{\beta, m_3} \hat \psi_{\alpha, m_4}
\end{align}
Here $\hat \psi_{\alpha, m}$ annihilates an electron in the zeroth Landau level (ZLL), with spin/valley quantum numbers $\alpha$, angular momentum $m$, and a spatial wavefunction with envelope $\psi_m(r)=\frac{1}{\sqrt{2^{m+1}\pi m!}} (\bar x+i\bar y)^m e^{-\frac{\bar x^2+\bar y^2}{4}}$, written in terms of dimensionless coordinates $\bar x=x/l_B$, $\bar y=y/l_B$ with magnetic length $l_B=\sqrt{\frac{h}{2\pi e B}}$. We introduce a spin- and valley-dependent chemical potential $\mu_\alpha$, which can account for the Zeeman effect and valley splittings due to the interaction between the graphene sample and the hBN substrate.  These terms can also be used to model aspects of the valley dependence of the interactions with electrons in the filled LL. We take the defect potential to be given by an attractive $+e$ Coulomb potential:
\begin{align}\label{impme}
    U_m &= 
    - \frac{e^2}{4\pi\epsilon \epsilon_0}\int \mathrm{d}^2r \frac{|\psi_m(r)|^2}{\sqrt{r^2+d^2}}
\end{align}
Here $d$ is the depth of the defect from the surface.
The Coulomb interactions between electrons have matrix elements:
\begin{align}
    V_{m_1 m_2 m_3 m_4} &= \frac{e^2}{4\pi\epsilon \epsilon_0}\int \frac{\mathrm{d}^2 r\, \mathrm{d}^2 r^\prime}{|r-r'|} \psi_{m_1}^*(r)\psi_{m_2}^*(r^\prime)\psi_{m_3}(r) \psi_{m_4}(r^\prime)
\end{align}
Our strategy for numerically calculating these matrix elements involving defect and electron interactions is given in Secs. \ref{subsubsec:coulomb_mat_defect} and \ref{subsubsec:coulomb_mat_electrons}.

We perform a particle-hole transformation for the electrons in one spin/valley state, denoted $K\!\!\uparrow$, corresponding to the filled LL.  We define $\hat \phi_m=\hat \psi_{K\uparrow, m}^\dagger$ and $\hat \phi_m^\dagger=\hat \psi_{K\uparrow, m}$, and take the index $\gamma$ (or $\gamma'$) to range over all other spin/valley quantum numbers $\gamma,\gamma^\prime\neq K\!\!\uparrow$.  After normal ordering,
\begin{align}\label{HH}
    H&=E_0+ \sum_m (-U_m-\bar{\mu}_h) \hat \phi_m^\dagger \hat \phi_m + \sum_{\gamma}(U_m-\bar{\mu}_\gamma) \hat \psi_{\gamma, m}^\dagger \hat \psi_{\gamma, m}+H_{hh}+H_{pp}+H_{ph}\\
    H_{hh}&= \frac{1}{2}\sum_{m_1m_2m_3m_4} V_{m_1m_2m_3m_4} \hat \phi_{m_1}^\dagger \hat \phi_{m_2}^\dagger \hat \phi_{m_3} \hat \phi_{m_4}\\
    H_{pp}&= \frac{1}{2}\sum_{\gamma \gamma^\prime}\sum_{m_1m_2m_3m_4} V_{m_1m_2m_3m_4} \hat \psi_{\gamma, m_1}^\dagger \hat \psi_{\gamma^\prime, m_2}^\dagger \hat \psi_{\gamma^\prime, m_3} \hat \psi_{\gamma, m_4}\\
    H_{ph}&=-\sum_{\gamma} \sum_{m_1m_2m_3m_4} V_{m_1m_2m_3m_4} \hat \phi_{m_4}^\dagger \hat \psi_{\gamma, m_2}^\dagger \hat \psi_{\gamma, m_3} \hat \phi_{m_1}
\end{align}
The terms $H_{hh}$, $H_{pp}$, and $H_{ph}$ are the Coulomb energies involving holes, electrons, and the interactions between them.
The energy $E_0=\sum_{mn} (V_{mnnm}-V_{mnmn})/2$ corresponds to the interactions between all of the electrons in the filled LL.  It is formally infinite, and can be understood as $E_0=
\frac{Q^2}{2C} - \frac{N E_x}{2}$, where $Q$ is the (infinite) total charge in the plane, $C$ is the capacitance, $N$ is the (infinite) number of electrons, and $E_x=\sum_m V_{mnmn}=\sqrt{\frac{\pi}{2}} E_c$ is the ``exchange energy" reflecting the fact that electrons are discrete and do not interact with themselves.   Here $E_c=\frac{e^2}{4\pi\epsilon \epsilon_0 l_B}$.  The chemical potentials are shifted by interactions with the filled states,
\begin{align}
    \bar{\mu}_h &= -\mu_{K\uparrow}+E_H-E_x \label{hole}\\
    \bar{\mu}_\gamma &= \mu_\gamma -E_H \label{particle}
\end{align}
The Hartree shift, $E_H=\sum_n V_{mnnm}=eQ/C$, is infinite and corresponds to the interaction of an electron with a uniform charge.  It can be absorbed into the chemical potentials.  Importantly, there is an exchange term in Eq. \ref{hole} but not \ref{particle}.

The Hamiltonian in Eq. \ref{HH} conserves the total number of particles of each flavor $N_\gamma =\sum_m \hat \psi_{\gamma, m}^\dagger \hat \psi_{\gamma, m}$, the total number of holes $N_h=\sum_m \hat \phi_m^\dagger \hat \phi_m$, and the total excess angular momentum, relative to the $\nu=-1$ ground state, $M=\sum_m m (\sum_{\gamma} \hat \psi_{\gamma, m}^\dagger \hat \psi_{\gamma, m}-\hat \phi_m^\dagger \hat \phi_m)$.  We can therefore label the many-body states with those quantum numbers.  In the absence of an impurity, the ground state is the $\nu=-1$ quantum Hall ferromagnet with $N_\gamma=N_h=0$, as long as the chemical potentials fall within the appropriate limits.  
If we neglect the Zeeman and valley splitting, so that each flavor of electron has the same chemical potential,
denoted $\mu=\mu_{K\uparrow} = \mu_\gamma$, then the energy to add a single hole or particle is $-\bar\mu_h$ or $-\bar \mu_\gamma$. These are positive if $E_H-E_x<\mu<E_H$.

In the presence of an attractive $+e$ impurity, we first consider the hole-free states where $N_h=0$, and a single extra electron in one of the other sectors.  These eigenstates  are of the form $\hat \psi_{\gamma, m}^\dagger |\Psi_{K \uparrow}\rangle$, where $|\Psi_{K \uparrow}\rangle$ is the vacuum of $\hat\psi_\gamma$ and $\hat\phi$, corresponding to filling the full $K\!\!\uparrow$ band. They have energy $U_m-\bar\mu_\gamma$; the lowest-energy state has $m=0$, so $E_{1e} = U_0 - \bar \mu_\gamma$ must be negative to make the electron binding energetically favorable. Again neglecting the Zeeman and valley splitting, we conclude that there will be an electron bound to the impurity if the common chemical potential obeys $\mu > U_0 + E_H$, where $U_0 = -\frac{e^2}{4\pi\epsilon \epsilon_0}\int \mathrm{d}^2r \frac{|\psi_0(r)|^2}{\sqrt{r^2+d^2}} = -E_c \sqrt{\frac{\pi}{2}} e^{\frac{d^2}{2 l_B^2}} \mathop{\rm erfc}\!\left(\frac{d}{\sqrt{2} l_B}\right)$.
There will be no electrons bound to the impurity if $E_H-E_x<\mu< U_0 + E_H$.  For a surface impurity, with $d=0$, this condition can never be satisfied, since $U_0(d=0)=E_x$.  However, if $d>0$, such a naked impurity is possible.

A surface/shallow impurity can also bind two electrons.
The two-electron states are of the form $|M;\gamma\gamma^\prime\rangle=
\sum_m c_m \hat \psi_{\gamma, m}^\dagger \hat \psi_{\gamma^\prime, M-m}^\dagger|\Psi_{K \uparrow}\rangle$.  For the special case where $M=0$ (which requires the two electrons to have different quantum numbers $\gamma\neq \gamma^\prime$ and $m=0$), the energy can again be calculated analytically, and 
\begin{align}
    E_{2e} &= (U_0 - \bar{\mu}_\gamma) + (U_0 - \bar{\mu}_{\gamma'}) + V_{0000}\\
    &=E_c \sqrt{\frac{\pi}{2}}\left(\frac{1}{\sqrt{2}}-2 e^{\frac{d^2}{2 l_B^2}} \mathop{\rm erfc}\!\left(\frac{d}{\sqrt{2} l_B}\right)\right) -\mu_\gamma-\mu_{\gamma'} + 2E_H
\end{align}
where we have used that the interaction energy between two electrons in $m=0$ state is $V_{0000}=\sqrt{\pi/4} E_c$.  Neglecting spin and valley splittings, the condition for the two-electron state to have a lower energy than the one-electron state, $E_{2e}<E_{1e}$, constrains the chemical potential to obey
$\mu>E_H+
 E_c \sqrt{\frac{\pi}{2}} \left(\frac{1}{\sqrt{2}}-e^{\frac{d^2}{2 l_B^2}} \mathop{\rm erfc}\!\left(\frac{d}{\sqrt{2} l_B}\right)
\right)$.
For stability of the $\nu=-1$ state, however, 
we also require that
 $\mu<E_H$, as otherwise it becomes favorable to add particles far from the impurity.  We conclude that a stable two-particle state may have a lower energy than the one-particle state only if $d$ is sufficiently small,
$d<d_{2e}$.  We numerically find that $d_{2e}\approx 0.48 l_B$. 
Deeper impurities are unable to bind more than one electron.


In addition to these charged configurations, we can consider spin-wave states.  These can be expressed as particle-hole pairs; for example, $|\Psi_{ph}\rangle=\sum_m c_m \hat\psi_{\gamma, m}^\dagger\hat \phi_{m-M}^\dagger |\Psi_{K\uparrow}\rangle$ corresponds to a spin wave with total angular momentum $M$.  In our finite numerics, we introduce a cutoff $m_{\rm max}$ and set $c_m=0$ for $m>m_\text{max}$.  Finding the spin-wave eigenstates then requires diagonalizing an $m_\text{max}\times m_\text{max}$ matrix.  While it is unlikely that the ground state of the system is described by one of these spin-wave states, as described below, they are created during a tunneling experiment.  One can also consider states with multiply flipped spins.

Another important class of states is the skyrmions, which contain one excess particle, $\sum_\gamma N_\gamma -N_h=1$, but $N_h\neq 0$.  Within a mean-field picture, they correspond to topologically nontrivial spin textures that cover the Bloch sphere once.  In the exact quantum state, the number of holes is related to the angular momentum by $M=N_h$ \cite{Abolfath1997}.  The Hilbert space grows as $m_\text{max}^{2N_h}$, and hence it can become very expensive to model these states for large $N_h$, which is related to the size of the skyrmion.  There are, however, a number of approximation schemes that have been developed (see, for example, \cite{Abolfath1997}). 
For impurities with sufficiently large $d$, a skyrmion state can be the ground state.  The size of the impurity, and hence the number of flipped spins $N_h$, grows with $d$.  The size is also sensitive to the spin and valley splittings, as well as the corrections to the electron-electron interaction that break the spin/valley symmetry \cite{Jolicoeur,PhysRevLett.117.056806,PhysRevB.95.245428}. Within the framework described here, the smallest skyrmions are described by states of the form $|\Psi_{pph}\rangle=\sum_{mn} c_{mn} \hat\psi_{\gamma, m}^\dagger\hat\psi_{\gamma^\prime, n}^\dagger\hat \phi_{m+n-M}^\dagger |\Psi_{K\uparrow}\rangle$.

To calculate the hole tunneling spectrum, we first choose the defect depth $d$ and each flavor chemical potential $\mu_\alpha$, and then find the ground state $|\Psi_0\rangle$, which has total angular momentum $M$ and energy $\epsilon_0$.  For each $m$, we then construct the Hilbert space that is accessible by tunneling a hole with quantum numbers $(\alpha, m)$ into the system.  For example, if we start with a one-electron bound state $|\Psi_0\rangle= \hat \psi_{K\downarrow, 0}^\dagger |\Psi_{K \uparrow}\rangle$ and take $\alpha=K\!\!\uparrow$, then the resulting space is spanned by the spin wave states with the basis states $\hat \phi_{m+n}^\dagger \hat \psi_{K\downarrow, n}^\dagger |\Psi_{K\uparrow}\rangle$.  
Conversely, if we take $\alpha=K\!\!\downarrow$ and further constrain $m=0$, we can produce the state $\hat\psi_{K\downarrow, 0} \hat\psi_{K\downarrow, 0}^\dagger |\Psi_{K\uparrow}\rangle = |\Psi_{K\uparrow}\rangle$.
We diagonalize the Hamiltonian in each of these sectors, finding the excited states $|x\rangle_{\alpha, m}$.  The STM tunneling hole/electron spectrum is proportional to the density of occupied/empty states, $\mathrm{d}I/\mathrm{d}V\propto A(r,\omega=eV_\text{B})$, where $A=A_h$ for $\omega<0$ and $A=A_p$ for $\omega>0$.  The hole tunneling spectrum is
\begin{align}\label{Ah}
    A_{h} (r, \omega) &= 
    \sum_{\alpha,x} \left|\langle x| \hat \psi_\alpha(r) |\Psi_0\rangle\right|^2\, 2\pi \delta(\epsilon_0-E_x-\omega)\\
    &=
    \sum_m |\psi_m(r)|^2 \sum_{\alpha, x} \left|\langle x| \hat \psi_{\alpha, m} |\Psi_0\rangle\right|^2\, 2\pi \delta(\epsilon_0-E_x-\omega)
\end{align}
where $r$ is the position of the tip, and $\hat \psi_{\alpha}(r)=\sum_m \psi_m(r) \hat \psi_{\alpha,m}$ is the operator that annihilates an electron at position $r$.
Electron tunneling is calculated similarly:
\begin{align}\label{Ap}
    A_{p} (r, \omega) = \sum_m |\psi_m(r)|^2 \sum_{\alpha, x} |\langle x| \hat \psi_{\alpha, m}^\dagger |\Psi_0\rangle|^2 2\pi \delta(E_x-\epsilon_0-\omega)
\end{align}
We choose chemical potentials $\mu_\alpha$ to match the experimental spectra, and broaden the delta functions into Lorentzians.

Fig. S\ref{fig:sim_v_-1} shows illustrative results, which demonstrate the sensitivity of the tunneling spectrum to the number of electrons bound to the defect.  
The experimental observations are consistent with a valley splitting of order 5-10 meV, and a smaller 
 Zeeman splitting (our parameter choices are given in the caption to Fig. S\ref{fig:sim_v_-1}).
 Fig. S\ref{fig:sim_v_-1}a, b, and c correspond to electron tunneling into empty states of a surface $+e$ defect at $\nu=-1$, with two (a), one (b), or zero (c) electrons bound to it.  At distances larger than a few magnetic lengths the broadened peaks from Eq. \ref{Ap} overlap, forming a smooth curve.  In Fig. S\ref{fig:sim_v_-1}a, where two electrons are bound to the impurity, this curve slopes downwards away from the origin, indicating that the tunneling-added electron is repelled by the net negative charge near the origin. Conversely, in the case of 1 or 0 bound electrons the slope is flat or upwards.  Similar features are seen for hole tunneling (Fig. S\ref{fig:sim_v_-1}d and e), and for a subsurface impurity (Fig. S\ref{fig:sim_v_-1}f to j).  In all cases of electron tunneling, two general separate branches are visible:  The weaker one comes from tunneling into the state with the same valley index as the filled LL ($K\!\!\uparrow$), but with opposite spin ($K\!\!\downarrow$).  The stronger line, with twice as much spectral weight, is associated with the two spin states that can tunnel into the other valley ($K^\prime\!\!\uparrow$, $K^\prime\!\!\downarrow$).  At $r=0$, when one or more electrons are bound to the defect, the electron excitation spectrum is dominated by a single line, as the probed empty states are all in the same valley $(K^\prime)$ with roughly the same energy.  Near $r\sim l_B$ the spectrum is complicated, and the excited states are superpositions of electrons in different $m$ modes. 

Fig. S\ref{fig:sim_v_-1}d and e instead show the hole tunneling spectrum, with either one or zero electrons bound to the defect.  In Fig. S\ref{fig:sim_v_-1}e we are tunneling a hole into a naked impurity, and the spectrum simply reflects the eigenstates $U_m$ of the defect potential.  Again, at large $r$ these broadened states overlap, forming a smooth curve whose slope indicates the charge of the screened impurity.  Unlike the electron excitation case, there is only one branch, as there is only one accessible flavor, $K\!\!\uparrow$.  At $r=0$, when tunneling into the state with one electron bound to the impurity (Fig. S\ref{fig:sim_v_-1}d), we observe one sharp peak and a broad continuum in the hole excitation spectrum.  The former corresponds to creating an eigenstate by removing the $K\!\!\downarrow$ electron which is bound to the impurity, as illustrated in Fig. S\ref{fig:sim_v_-1}k to l. Conversely, the state $|{\rm SW}\rangle_0$ formed by removing a $K\!\!\uparrow$ from the $m=0$ state of the filled LL is not an eigenstate -- the Coulomb interaction between electrons couples all the $M=0$ states with one flipped spin, $|{\rm SW}\rangle_n=\hat \psi_{K\uparrow,n}\hat \psi_{K\downarrow,n}^\dagger |\Psi_{K,\uparrow}\rangle$ (Fig. S\ref{fig:sim_v_-1}m).  As already argued, the energy eigenstates are spin waves, corresponding to superpositions of all these states.

The spectra are similar for a subsurface defect (Fig. S\ref{fig:sim_v_-1}f to j).  The most notable difference is that due to a weaker defect potential, the state at $m=0$ tends to be repelled up to higher energies under overscreening by bound electrons.

If one integrates Eq.~\ref{Ah} over energy, one can derive the sum rule:
\begin{align}
    \int \frac{\mathrm{d}\omega}{2\pi}\,A_h(r,\omega) 
    &= \sum_{\alpha,x} \left|\langle x| \hat \psi_\alpha(r) |\Psi_0\rangle\right|^2\\
    &= \sum_\alpha \langle \Psi_0| \hat \psi_\alpha^\dagger (r) \hat \psi_\alpha(r) |\Psi_0\rangle
\end{align}
The integrated hole spectral density gives the local density of occupied states.  Similarly, 
\begin{align}
    \int \frac{\mathrm{d} \omega}{2\pi}\,A_p(r,\omega) 
    &=\sum_\alpha \langle \Psi_0| \hat \psi_\alpha(r) \hat \psi_\alpha^\dagger (r) |\Psi_0\rangle
\end{align}
If the spectral features from different spins/valleys are well separated, one can distinguish their contributions.

Fig. S\ref{fig:s6} shows an example of tunneling a hole into a surface $+e$ defect with one bound electron.  When we exclude the spin wave continuum and include the bound electron, the spectral weight integral above the dashed red line reveals a flat density (Fig. S\ref{fig:s6}b), while the total spectral weight gives the occupied $K\!\!\uparrow$ band with an additional bound electron density (Fig. S\ref{fig:s6}c).  The experimental spectrum is somewhat more complicated, displaying a series of dispersing peaks rather than a smooth continuum.  This possibly indicates the presence of standing spin waves.  As we discuss Sec. \ref{subsubsec:tip-sim}, the cusps near $r=0$ are due to screening of the impurity by the STM tip.

\subsubsection{Simulation of tip effects}
\label{subsubsec:tip-sim}

We can extend our modeling to include the electrostatic contributions from the tip. The charge on the tip creates a potential that influences the electronic states. Additionally, the metallic tip screens the interactions both between nearby electrons and between the electrons and the impurity. Because we consider the tunneling of a single hole into a LL near an impurity, electron-electron interactions play no role beyond providing a uniform shift to all energies. Thus, we only need to model the charging effects and the screening of the impurity.

As illustrated in Fig. S\ref{fig:tip_sim}, we model the tip as a metallic sphere of radius $R$, centered a distance $H$ above the graphene surface. By choosing the origin on the graphene surface immediately below the tip, we locate the tip center at coordinates $(0,0,H)$. The impurity, with charge $+e$, is located at $(x,0,-d)$, where $x$ is the horizontal distance from the tip to the impurity and $d$ is its depth below the surface. To account for screening by the tip, we place an image charge $Q = -\frac{R}{r}e$ at a distance $s=\frac{R^2}{r}$ from the sphere's center along the line connecting the center to the impurity, where $r=\sqrt{x^2+(H+d)^2}$ is the distance between them. The coordinates of the image charge are therefore $(x_\text{image}, y_\text{image}, z_\text{image})=\left(\frac{s}{r}x, 0, H-\frac{s}{r}(H+d)\right)$. We then solve the single-particle problem for an electron in the LL subject to the total potential energy $U(\bar x,\bar y)=U_\text{imp}(\bar x,\bar y)+U_\text{image}(\bar x,\bar y) + U_\text{tip}(\bar x,\bar y)$, given by:
\begin{align}
    U_\text{imp}(\bar x, \bar y) &= \frac{-e^2}{4\pi\epsilon\epsilon_0} \frac{1}{\sqrt{(\bar x-x)^2+\bar y^2+d^2}}\\
    U_\text{image}(\bar x, \bar y) &= \frac{-e Q}{4\pi\epsilon\epsilon_0}\frac{1}{\sqrt{(\bar x-x_\text{image})^2+\bar y^2+z_\text{image}^2}}\\
    U_\text{tip}(\bar x, \bar y) &= \frac{-e}{4\pi \epsilon \epsilon_0} \frac{-4\pi \epsilon \epsilon_0 R V_t}{\sqrt{\bar x^2+\bar y^2 + H^2}} = \frac{e R V_t}{\sqrt{\bar x^2+\bar y^2+H^2}}.
\end{align}
Here, $V_t = V_\text{B}-W$ is the effective tip voltage, related to the tip-sample bias voltage $V_\text{B}$ and the tip-to-sample work function difference $W$. Thus, $U_\text{tip}$ roughly accounts for the charge accumulated on the tip-sample capacitor. We construct the single-particle Hamiltonian matrix $H_{mn}(x)$, where 
\begin{align}\label{Hmn}
    H_{mn}(x) = \int \mathrm{d}\bar x \, \mathrm{d}\bar y \, \psi_m^*(\bar x, \bar y) U(\bar x, \bar y) \psi_n(\bar x, \bar y).
\end{align}
Here, $\psi_m$ is the LL wavefunction with angular momentum $m$. We choose a cutoff $m_\text{max}$ and truncate $H$ to an $m_\text{max}\times m_\text{max}$ matrix. Numerical details are provided in subsequent Sec. \ref{subsubsec:tip-matrixelements}. For each tip-impurity separation $x$, we numerically diagonalize this matrix, finding the eigenvalues and eigenvectors satisfying
\begin{align}\label{eigen}
    \sum_{m^\prime} H_{mm^\prime}(x) \chi_{m^\prime}^{(n)} = E_n(x) \chi_{m}^{(n)}.
\end{align}

In our simplest modeling, we treat $V_t$ as a constant (which depends on the intrinsic charge on the tip, and includes the image charge from the graphene electrons). A more sophisticated model involves recognizing that $V_t$ varies linearly with the bias voltage, which in turn is proportional to $E_n$ when one is tunneling in holes with that energy. In this second model, we set $eV_t = E_n - eW$, in which case Eq. \ref{eigen} transforms into a generalized eigenvalue problem. We find nearly identical results when using this energy-dependent form.  In the subsequent discussion, we use the simpler model, where $V_t$ is chosen to be a constant.

Given that we use a spatial basis centered at the location of the tip, a tunneling hole couples exclusively to the $m=0$ state, which is localized below the tip. We therefore calculate the tunneling spectrum as
\begin{equation}
    A(x, E) \propto \sum_n |\chi_0^{(n)}(x)|^2 \delta(E - E_n(x)).
\end{equation}

Figs. S\ref{fig:tip_sim}d and S\ref{fig:tip_sim}f compare this model with the experiment over the ZLL of a surface $+e$ defect. In Figs. S\ref{fig:tip_sim}b and S\ref{fig:tip_sim}c, we repeat the calculation using the wavefunctions for the N = -1 LL to explore the tunneling of a hole into the $\nu=-2$ state. 
A sharp feature observed near $x=0$ is attributed to tip screening of the impurity. Due to our crude modeling of the tip as a sphere, this feature is somewhat more rounded in the theoretical calculation than in the experiment. Additionally, each spectral segment appears concave and disperses with the tip position. This feature is driven by $U_\text{tip}$ and can be readily understood perturbatively. To zeroth order, the eigenstates are simply the symmetric-gauge Landau wavefunctions $\psi_m$ centered at the impurity, with energies $E_m^{(0)}$. The first-order energy correction is $E_m^{(1)}(x)= \int \mathrm{d}^2 \bar r\, U_\text{tip}(\bar r) |\psi_m(\bar r)|^2$. This correction is maximized when the tip sits directly above the density maximum of the wavefunction and diminishes as the tip is moved to either side, resulting in the observed cupped shape. The direction of this curvature depends on the sign of $V_t$, which is intrinsically tied to the tip-sample work function difference. 
In Fig. S\ref{fig:tip_sim}b, the magnitude of $V_\text{B}$ is approximately three times larger than in Fig. S\ref{fig:tip_sim}d, resulting in much stronger tip-induced spectral bending.

Conversely, when $V_t$ is sufficiently large (or the subsurface defect is sufficiently weak), one can instead treat the defect as a small perturbation. In this regime, the tunneling spectrum displays a single peak that continuously disperses with the tip position. 
The experimental examples shown here lie in an intermediate regime.
Similar modeling can also be carried out for the $\nu=\pm 1$ states. In those cases, however, $V_t$ should be much smaller in magnitude if an appropriate tip is chosen. Consequently, the most prominent electrostatic effect becomes the screening peak at small tip-impurity separations $r$. The data in Fig. S\ref{fig:s4} provide an illustrative example of this behavior.

Importantly, these electrostatic tip effects could be used to manipulate quasiparticles in the sample. For example, one could trap an excitation within the tip's potential and physically move it around as part of a braiding experiment.

\subsubsection{Calculating displaced Coulomb matrix elements}
\label{subsubsec:tip-matrixelements}

Here we describe our numerical strategy for calculating the matrix elements in Eq. \ref{Hmn}. We begin by constructing the magnetic translation operator $\hat T_w$, which transforms ZLL wavefunctions as
\begin{equation}\label{mag}
    \hat T_w[\psi](z) = \psi(z-w) e^{(z w^*-z^* w)/4}
\end{equation}
Here, $z=(x+iy)/l_B$ is a dimensionless complex number representing the position in the plane, and $\psi(z)=f(z) e^{-|z|^2/4}$, where $f$ is an analytic function. The exponential factor in Eq. \ref{mag} is a phase required to preserve this analytic form. We consider the case where $w=x$ is real, corresponding to a translation along the $x$-axis. We construct the operator $\hat T_x$ by exponentiating the generator of infinitesimal magnetic translations, $\hat T_x=e^{x \hat D}$. The generator $\hat D[\psi](z)=(-\partial_x+iy/2) \psi(z)$ has matrix elements $D_{mn}=\langle m| \hat D |n\rangle$ given by
\begin{equation}
    D_{mn} = \sqrt{\frac{m}{2}} \delta_{n,m-1} - \sqrt{\frac{n}{2}}\delta_{n,m+1}
\end{equation}
We truncate the basis to $m<m_\text{max}$ and construct $D$ as an $m_\text{max}\times m_\text{max}$ matrix. We then matrix-exponentiate this to produce $T_x$. The matrix elements for a shifted Coulomb potential are then evaluated via the similarity transformation
\begin{equation}\label{shift}
    U^x_{mn} = \sum_{m^\prime} (T_x)_{mm^\prime}  U_{m^\prime} (T_x^\dagger)_{m^\prime n}
\end{equation}
where $U_{m^\prime}$ are the diagonal matrix elements of the unshifted potential given by Eq. \ref{impme}. Finally, we construct the total single-particle Hamiltonian $H_{mn}$ by summing the appropriately shifted contributions from the impurity, the image charge, and the tip. The impurity and image charge contributions are both calculated by using Eq. \ref{shift}, but with different $U_m$'s and $T_x$'s, reflecting the different distances of the charges from the surface, and different horizontal displacements of the the charges from the tip location. Calculations for higher LLs can be performed similarly, as their wavefunctions can be obtained recursively from the ZLL wavefunctions.

\subsubsection{Simulation of spectra of the \texorpdfstring{$\bm \nu^*$}{v*} = 1/3 fractional quantum Hall state}
\label{subsubsec:theory_v1/3}

The fractional quantum Hall states that we study are fully polarized, and it suffices to consider spinless particles at filling $\nu^*=1/3$. We place these particles on a spherical shell of radius $R$ enclosing a magnetic monopole, such that exactly $2Q$ magnetic flux quanta thread the surface of the sphere. The lowest LL wavefunctions are \cite{Haldane1985,Arciniaga2016}
\begin{align}
    \psi_{Qm}(\theta,\phi) &=
    \left[
    \frac{2Q+1}{4\pi} \binom{2Q}{Q-m}
    \right]^{1/2}
    (-1)^{Q-m} u^{Q+m} v^{Q-m}
\end{align}
where $u = \cos(\theta/2) e^{i\phi/2}$, $v = \sin(\theta/2) e^{-i\phi/2}$, and $m$ takes the $2Q+1$ values from $-Q$ to $Q$. The $\nu^*=1/3$ Laughlin state corresponds to the relation $N = 2Q/3 + 1$ \cite{Haldane1985}; a mismatch of $Q$ away from this value for a given $N$ introduces extra quasiparticles or quasiholes, each of which carries charge $\pm e/3$. This boundary-free geometry is chosen to minimize finite-size effects.

We define the Coulomb matrix elements,
\begin{align}
    V_{m_1m_2m_3m_4} &= \frac{e^2}{4\pi \epsilon \epsilon_0}\int \mathrm{d}\theta\,\mathrm{d}\theta^\prime\,\mathrm{d}\phi\,\mathrm{d}\phi^\prime\, \frac{\sin \theta \sin \theta^\prime}{|\vec r-\vec r^\prime|}
    \psi_{Q m_1}^*(\vec r) \psi_{Q m_2}^*(\vec{r}^\prime) \psi_{Q m_3}(\vec{r}^\prime) \psi_{Q m_4}(\vec{r})
\end{align}
which vanish unless $m_1+m_2=m_3+m_4$.  Our approach to calculating it is given in Sec. \ref{subsubsec:trick_fractional}.  We also define the interaction matrix elements between the electrons and a $+e$ Coulomb impurity, placed at a distance $d$ above the north pole of the sphere,
\begin{align}
    V_m &= -\frac{e^2}{4\pi \epsilon \epsilon_0}\int \mathrm{d}\theta\,\mathrm{d}\phi\,\frac{\sin \theta}{|\vec{r}-(R+d)\hat z|} |\psi_{Qm}(\theta,\phi)|^2
\end{align}
The many-body Hamiltonian is then expressed as
\begin{align}\label{ham}
    H &= \sum_m V_m \hat \psi_m^\dagger \hat \psi_m + \frac{1}{2}\sum_{m_1 m_2 m_3 m_4} V_{m_1m_2m_3m_4} \hat \psi_{m_1}^\dagger \hat \psi_{m_2}^\dagger \hat \psi_{m_3} \hat \psi_{m_4}
\end{align}

We use a matrix product state (MPS) ansatz to approximate the ground state of this Hamiltonian, writing \cite{Schollwck2011}
\begin{align}
|\Psi\rangle = \sum_{\{s_m\}}\prod_m |A^{(m)}_{s_m s_{m+1}}\rangle
\end{align}
where for each pair of link indices $s_m$ and $s_{m+1}$, $|A^{(m)}_{s_m s_{m+1}}\rangle=A^{(m)}_{s_m s_{m+1}, 0} |0\rangle_m + A^{(m)}_{s_m s_{m+1}, 1} |1\rangle_m$ is a state vector in the local Hilbert space of the mode with angular momentum $m$. Mode $m$ is occupied/empty in states $|1\rangle_m$/$|0\rangle_m$.  The link indices $s_m$ take up to $\chi$ different values, a number referred to as the bond dimension. This form of the wavefunction is highly expressive, and the ansatz becomes exact when $\chi$ is made sufficiently large. It provides an efficient representation of the wavefunction when the entanglement between the various modes is small. Variants of this approach have been used to study the fractional quantum Hall effect in various geometries \cite{Liu2015,Zaletel2015,HU20122157,Shibata2001,Feiguin2008,Zhao2011,Motruk2016,Misguich2021}.

We use the density matrix renormalization group (DMRG) algorithm to optimize the matrices $|A^{(m)}_{s_m s_{m+1}}\rangle$ and find the ground state $|\Psi_0\rangle$ of minimal energy.  The Hamiltonian in Eq. \ref{ham} is written as a matrix product operator (MPO).  Due to the long-range nature of the Coulomb interaction, this operator has a large bond dimension.  We use an optimal compression scheme based on sequential singular value decompositions.  By repeating our calculations with several values of our cutoff, we verify that our results are not affected by the resulting truncation error, nor by the truncation error in the MPS wavefunction.

To calculate the STM spectrum, we use a global Krylov method to calculate the relevant excitations, similar to \cite{PhysRevB.85.205119,PhysRevB.60.335,PhysRevE.94.053308}.  Complementary approaches are based upon Chebychev moments \cite{PhysRevB.83.195115,PhysRevB.90.115124,PhysRevB.90.045144,PhysRevB.91.115144}.  Starting with the state $|\Phi_0\rangle=\hat\psi_m |\Psi_0\rangle$, which corresponds to the creation of a hole with angular momentum $m$ in the tunneling process, we generate an orthogonal basis $\{|\Phi_n\rangle\}$ for the Krylov space spanned by $\{H^n |\Phi_0\rangle\}$ for $n=0,1,2,\cdots n_\text{max}$.  We express this bundle of states as a multi-targeted matrix product state, in mixed canonical form, 
\begin{align}
    |\Phi_n\rangle &=
    \sum_{\{s_m\}} \left(\prod_{l<c} |L^{l}_{s_l s_{l+1}}\rangle\right)
    |C^{c, n}_{s_{c} s_{c+1}}\rangle
    \left(
    \prod_{r>c}
    |R^{r}_{s_r s_{r+1}}\rangle
    \right)
\end{align}
Here $c$ is the angular momentum corresponding to the ``orthogonality center'', and $l,r$ are summed over the angular momenta which are smaller than or greater than this value.  The left and right matrices are isometries,
\begin{align}\label{L}
    \sum_{s} \langle L^{l}_{s t^\prime}| L^{l}_{s t}\rangle &= \delta_{t^\prime,t}\\
    \sum_{s} \langle R^{r}_{ t^\prime s}| R^{r}_{ts}\rangle &= \delta_{t^\prime,t}
    \label{R}
\end{align}
and represent a common basis for the states in the Krylov space.  The only matrix in the decomposition which depends on the state index $n$ is $|C^{c, n}_{s_{c} s_{c+1}}\rangle$.  This common basis minimizes the truncation error which occurs when we need to take linear superpositions of the states.  Due to the properties in Eq. \ref{L} and \ref{R}, we can efficiently calculate the inner products,
\begin{align}
    \langle \Phi_i |\Phi_j\rangle = \sum_{s t} \langle C^{c,i}_{st}| C^{c,j}_{st}\rangle
\end{align}

We use a variant of the Lanczos algorithm to generate an orthogonal basis for the Krylov space \cite{lanczos}.  Since $H$ is Hermitian, one can iteratively build the basis vectors via
\begin{align}\label{lanczos}
    |\Phi_n \rangle &\propto H |\Phi_{n-1}\rangle - a_{n-1} |\Phi_{n-1} \rangle - b_{n-1} |\Phi_{n-2}\rangle
\end{align}
where $a_n = \langle \Phi_n|H|\Phi_n\rangle$ and $b_n = \langle \Phi_{n-1}|H|\Phi_n\rangle$.  The Hamiltonian, when projected onto this basis, is a tridiagonal matrix with the $a_n$ on the diagonal and the $b_n$ on the off‑diagonals.

Each time we expand the Krylov space, we recompute the rightmost isometries $|R^{Q}_{s}\rangle$ as the eigenvectors of the reduced density matrix
\begin{align}
    \rho^Q &= \operatorname{Tr}_{m<Q} \sum_n |\Phi_n\rangle \langle \Phi_n| \nonumber\\
    &= \sum_s \lambda_s |R^{Q}_s\rangle \langle R^{Q}_s|
\end{align}
where the trace is taken over all modes with $m<Q$.  Thus $\rho^Q$ is a $d\times d$ matrix, and $|R_s^Q\rangle$ is a length‑$d$ vector, with $d=2$ the dimension of the local Hilbert space associated with the $m=Q$ states.
This construction lets us write each wavefunction as
\begin{align}
    |\Phi_n\rangle &= \sum_s |\Phi_{ns}^{(Q-1)}\rangle\, |R^{Q}_s \rangle
\end{align}
where $|\Phi_{ns}^{(Q-1)}\rangle = \langle R^{Q}_s|\Phi_n\rangle$ is a wavefunction living on the modes with $m<Q$.  We then construct the next isometry by decomposing the density matrix
\begin{align}
    \rho^{Q-1}_{ss'} &= \operatorname{Tr}_{m<Q-1} \sum_n 
    |\Phi_{ns}^{(Q-1)}\rangle \langle \Phi_{ns'}^{(Q-1)}| \nonumber\\
    &= \sum_t \lambda_t |R^{(Q-1)}_{ts}\rangle \langle R^{(Q-1)}_{ts}|
\end{align}
This procedure is repeated until we reach $m=-Q$, yielding the whole bundle of states in the right‑canonical form.  At each step we truncate the bond dimension by discarding density‑matrix eigenvalues $\lambda$ that fall below a chosen threshold.  Because of this truncation, the resulting basis is no longer strictly orthogonal.  We therefore apply the Gram–Schmidt algorithm to the $C^{(1)}$ tensors to reorthogonalize the states, updating them without altering the $R$ tensors.  To account for round‑off errors, the Gram–Schmidt procedure is iterated until convergence.

Both round-off and truncation effects introduce small errors in the wavefunctions, $|\Phi_n^\text{approx}\rangle = |\Phi_n^\text{exact}\rangle + |\delta_n\rangle$, where $|\delta_n\rangle$ lies outside the Krylov space.  When one uses variants of this algorithm to find eigenvalues of a matrix, these errors can be useful because they prevent the Krylov space from being exhausted, and are key to the strategy of \emph{implicit restart}.  In the context of our spectral problem, however, they are a nuisance and lead to spurious ``ghost'' eigenvalues.  We eliminate these using a deflation technique.  
We identify spurious eigenvectors by diagonalizing the Hamiltonian projected onto the approximate Krylov space (often referred to as the Rayleigh quotient).  Any eigenvector whose overlap with $|\Phi_0\rangle$ falls below some threshold is deemed spurious.  We eliminate it by using a shifted QR transformation.

We continue to grow the Krylov space, alternating the Lanczos construction (Eq. \ref{lanczos}) with tests for spurious states.  We continue until all non-spurious eigenstates are accurately captured within the truncated Krylov space to within a chosen precision.  We then use this basis to approximate
\begin{align}
    A_h^m(\omega) &=\sum_f |\langle f| \hat \psi_m |\Psi_0\rangle|^2 \, 2\pi\delta\bigl(\omega-(E_f-E_0)\bigr) \label{Ah_f}\\
    &=\sum_f |\langle  f|\Phi_0 \rangle|^2 \, 2\pi\delta\bigl(\omega-(E_f-E_0)\bigr)
\end{align}
which we evaluate using the eigenvectors of the projected Hamiltonian.  From these functions we can extract
\begin{align}
    A_h (\omega,r)=\sum_m |\psi_{Qm}(r)|^2 A_h^m (\omega) \label{Ahr}
\end{align}

To verify these results, we use an alternative approach that constructs excited states sequentially.  Starting with the ground state $|\Psi_0\rangle$, we find the first excited state as the state corresponding to the smallest eigenvalue of $H_1 = H + \lambda\,|\Psi_0\rangle\langle\Psi_0|$, where $\lambda$ is a penalty coefficient, chosen such that $\lambda > E_1 - E_0$.  The second excited state is the state associated with the smallest eigenvalue of $H_2 = H_1 + \lambda\,|\Psi_1\rangle\langle\Psi_1|$, and so on.  We continue until the retained states capture the bulk of the spectral weight,
\begin{align}
\langle \Psi_0|\hat \psi_m^\dagger \hat \psi_m|\Psi_0\rangle - \sum_n 
\langle \Psi_0|\hat \psi_m^\dagger|\Psi_n\rangle \langle \Psi_n| \hat \psi_m|\Psi_0\rangle < \epsilon
\end{align}
where $\epsilon$ is a small tolerance.  Once the relevant excited states have been found, we calculate the tunneling spectrum using Eqs. \ref{Ah_f} and \ref{Ahr}.  We find that the two approaches yield identical results, though the Lanczos approach requires many fewer numerical resources.

For all of these calculations, we make use of the ITensor Software Library \cite{itensor}, and we work with tensors where the number of particles and the angular momentum are good quantum numbers.

Fig. S\ref{fig:fractional_sim} shows representative examples in which the defect depth $d$ is chosen such that the $\mp e$ impurity binds different numbers of charge $\pm e/3$ quasiparticles. In the numerical calculations, the number of quasiparticles manually introduced in the initial DMRG state is chosen to match the number of quasiparticles bound to the defect. For a $-e$ defect (Fig. S\ref{fig:fractional_sim}a-d), increasing the number of bound quasiparticles pushes up the energy of the $m=0$ state and modifies the spectral features at short distances.  For a $+e$ defect, one has no spectral weight at $r=0$ once a single quasihole is bound -- and the size of that spectral hole is indicative of the number of quasiholes.  Note, however, that the qualitative features of the spectra with different numbers of quasiparticles are quite similar to one another.  A quantitative comparison is needed to determine the amount of bound charge. 

Other authors have argued that the multiplicity of modes can be used as a signature of quasiparticles -- but that approach requires detecting spectral features with extremely small spectral weight \cite{Papic2018}.  Additionally, there are arguments relating the tunneling spectrum far from the impurity to composite fermions \cite{Gattu2024,Pu2024,Yue2024}.

\subsubsection{Numerically calculating Coulomb matrix elements with defect}
\label{subsubsec:coulomb_mat_defect}

To calculate the matrix elements in Eq. \ref{impme}, we define 
\begin{align}\label{WM}
    W_m &= \int \mathrm{d} r \frac{r^{2m+1}}{\sqrt{r^2+d_B^2}} e^{-r^2/2}
\end{align}
in terms of which $U_m= - \frac{1}{2^m m!} W_m E_c$, where $E_c=\frac{e^2}{4\pi\epsilon \epsilon_0 l_B}$ has already been introduced as the characteristic Coulomb energy in the LL.  The position vector in Eq. \ref{WM} is dimensionless, representing the position in units of the magnetic length, and $d_B=d/l_B$ is the defect depth in those units.  To calculate $W_m$, we introduce the generating function 
\begin{align}
    F(s)&=\sum_{m=0}^\infty \frac{s^m}{m!} W_m=\frac{\sqrt{\pi}}{2\sqrt{1/2-s}} e^{(1/2-s) d_B^2} \mathop{\rm erfc}(d_B\sqrt{1/2-s})
\end{align}
where ${\rm erfc}(x)$ is the complementary error function.  To extract the coefficients,
we differentiate $F(s)$ with respect to $s$. After integrating by parts, we find
\begin{align}
    (1-2 s) F'(s) &= d_B + (1-d_B^2+ 2 d_B^2 s) F(s)
\end{align}
which corresponds to the recursion relationship
\begin{align}
    W_{m+1} &= (1-d_B^2+2m) W_m + 2 d_B^2 m W_{m-1} \quad (m\geq 1)
\end{align}
which can be used to efficiently calculate the coefficients, as long as $d_B$ is not too large.  For large $d_B$, the recursion relation is unstable, and instead we simply numerically calculate the integral in Eq. \ref{WM} using Gaussian quadrature.

\subsubsection{Numerically calculating Coulomb matrix elements between electrons}
\label{subsubsec:coulomb_mat_electrons}

To calculate the interactions between electrons, we write 
\begin{align}
V_{m_1m_2m_3m_4} &= E_c \frac{2^{-(m_1+m_2+m_3+m_4+4)/2}}{\pi^2\sqrt{m_1! m_2! m_3! m_4!}} X_{m_1m_2m_3m_4}\\
X_{m_1m_2m_3m_4}&=\int \frac{\mathrm{d}^2 r\,\mathrm{d}^{2} r^\prime}{|r-r^\prime|} (x-iy)^{m_1}(x^\prime-iy^\prime)^{m_2} (x+iy)^{m_3} (x'+iy')^{m_4} e^{-\frac{x^2+y^2+(x^\prime)^2+(y^\prime)^2)}{2}}
\end{align}
where, as already introduced, $E_c=\frac{e^2}{4\pi\epsilon \epsilon_0 l_B}$, and the lengths in this expression are dimensionless.  We transform to center-of-mass and relative coordinates, defining 
$(x+iy)=(X+iY)+(u+iv)/2$ and $(x^\prime+i y^\prime)=(X+i Y)-(u+iv)/2$.  We use the binomial theorem to expand terms such as $(x-iy)^{m_1}$ as the sum of terms involving powers of the center of mass and relative coordinates, finding
\begin{align}
    X_{m_1m_2m_3m_4}&=2\pi^2 \delta_{m_1+m_2-m_3-m_4}\sum_{s=0}^{m_1 + m_2} \Gamma(s+\frac{1}{2}) (m-s)! \Lambda_{m_1 m_2}^{(s)} \Lambda_{m_3m_4}^{(s)}\\
    \Lambda_{mn}^{(s)}&=
    \sum_{\bar{s}=0}^s {{m}\choose{\bar s}}{{n}\choose{s-\bar s}}  (-1)^{\bar s}
\end{align}
We can efficiently calculate the coefficients $\Lambda_{m_1m_2}^{(s)}$ by introducing a generating function
\begin{align}
G_{mn}(t) &= \sum_{s=0}^\infty \Lambda_{mn}^{(s)} t^s=(1-t)^{m} (1+t)^n
\end{align}
Noting that $(1-t) G_{mn}(t)=G_{m+1,n}(t)$, we obtain a recursion relation
\begin{align}
\Lambda_{m+1,n}^{(s)}&=
\Lambda_{mn}^{(s)} - \Lambda_{mn}^{(s-1)} \quad (s \geq 1)
\end{align}
that can be used to efficiently tabulate the $\Lambda$'s.  We store these and use them to rapidly calculate the Coulomb matrix elements.

\subsubsection{Calculating Coulomb matrix elements on a sphere}
\label{subsubsec:trick_fractional}

We use the expression from \cite{Arciniaga2016}, 
\begin{align}
    V_{m_1m_2m_3m_4} &= \frac{e^2}{4\pi \epsilon \epsilon_0}\int \mathrm{d}\theta\,\mathrm{d}\theta^\prime\,\mathrm{d}\phi\,\mathrm{d}\phi^\prime\, \frac{\sin(\theta)\sin(\theta^\prime)}{|\vec r-\vec r^\prime|}
    \phi_{m_1}^*(\vec r) \phi_{m_2}^*(\vec{r}^\prime) \phi_{m_3}(\vec{r}^\prime) \phi_{m_4}(\vec{r})\\
    &=\frac{e^2}{4\pi\epsilon \epsilon_0 R}(2Q+1)^2(-1)^{m_3+m_4}
    \sum_{l_0=0}^{2 Q}\sum_{m=-l_0}^{l_0} (-1)^m
    \\&\quad\times
    \left\{\begin{array}{ccc}
        Q&l_0&Q\\
        m_4&m&-m_2
    \end{array}
    \right\}
    \left\{\begin{array}{ccc}
        Q&l_0&Q\\
        m_3&-m&-m_1
    \end{array}
    \right\}
    \left(
    \left\{\begin{array}{ccc}
        Q&l_0&Q\\
        -Q&0&Q
    \end{array}
    \right\}
    \right)^2
\end{align}
where the symbols shown are Wigner three-$j$ symbols, which vanish unless the sum of the three lower elements is zero.  Consequently, $V$ vanishes unless $m_1+m_2=m_3+m_4$, and the sum over $m$ reduces to the single term $m=m_2-m_4=m_3-m_1$.  The expression simplifies further using the identity
\begin{align}
    \left(\left\{\begin{array}{ccc}
    Q&l_0&Q\\
    -Q&0&Q
    \end{array}
    \right\}\right)^2=\frac{(2Q)!(2Q)!}{(2Q-l_0)!(2Q+l_0+1)!}.
\end{align}

For the impurity potential $V_m$, we evaluate the matrix elements numerically via Gaussian quadrature.  For the special case of a surface impurity (the impurity placed directly on the sphere, $d=0$), the integral can be performed analytically, yielding
\begin{align}
    V_m&= -\frac{e^2}{4\pi\epsilon_0\epsilon R} \frac{1}{2} 
    \frac{\Gamma(Q-m+1/2)}{\Gamma(Q-m+1)} \frac{\Gamma(2Q+2)}{\Gamma(2Q+3/2)}.
\end{align}

\clearpage



\phantomsection
\addcontentsline{toc}{subsection}{Fig. S1 \textbar{} Device images and gate-dependent d\textit{I}/d\textit{V} spectra.}

\begin{figure*}
    \centering
    \refstepcounter{figure}
    \label{fig:device_spectra}
    \includegraphics[width=\linewidth]{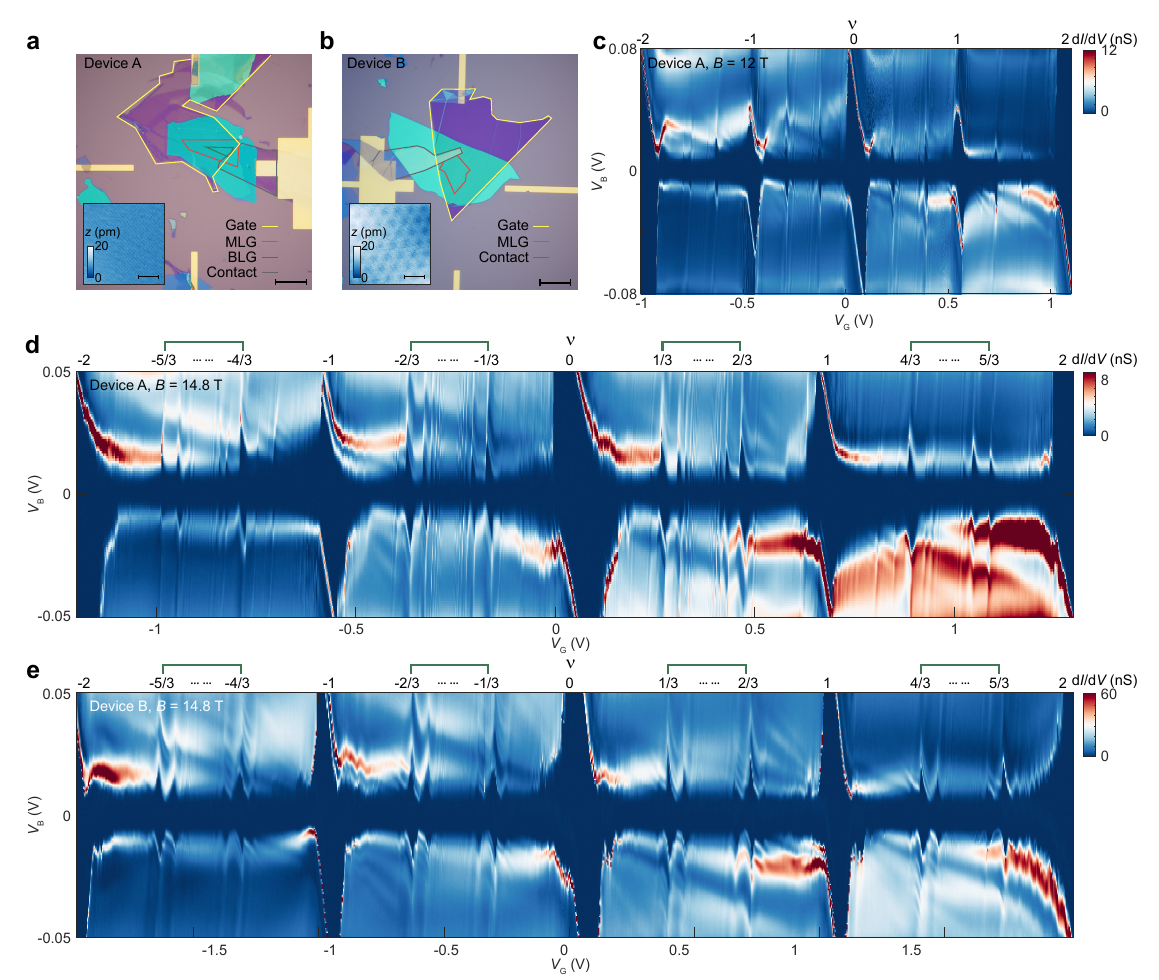}
\end{figure*}
\noindent\textbf{Fig. S1 \textbar{} Device images and gate-dependent d\textit{I}/d\textit{V} spectra.}
\textbf{a}, \textbf{b}, Optical images of devices A and B. Graphene and bottom graphite gate are connected to pre-patterned electrodes. Graphite gate, monolayer graphene (MLG), bilayer graphene (BLG), and contact graphite are outlined by yellow, red, green, and gray lines, respectively. Scale bars: 20 $\mu$m. Insets show their typical STM images acquired at 4.2 K (scale bars: 2 nm). We observe no mori\'e pattern in device A, while device B shows a mori\'e pattern with a 1.2 nm period formed with the hBN substrate. The STM images in the insets of \textbf{a} and \textbf{b} are acquired at $V_\text{B}$ = 200 mV, $I_\text{t}$ = 100 pA and $V_\text{B}$ = 400 mV, $I_\text{t}$ = 10 pA, respectively. In this work, all integer-filling-related data are acquired on device A while $\nu = \pm 5/3$ data are acquired on device B, unless otherwise specified in the figure caption. This is because $\nu = \pm 5/3$ gap is much more pronounced for devices with better alignment with hBN.
\textbf{c}, \textbf{d}, \textbf{e},  Constant height gate-dependent d$I$/d$V$ spectra at 12 T and 14.8 T measured with constant-height-setpoint method (see Methods). These data exhibit Haldane sashes, indicating that the tip-gating effect is not pronounced \cite{farahiBrokenSymmetriesExcitation2023a}. Jain sequence fractional quantum Hall gaps (marked by green solid lines) are also evident, demonstrating the high quality of our devices.
\clearpage

\phantomsection
\addcontentsline{toc}{subsection}{Fig. S2 \textbar{} Comparison between surface topography and simultaneously acquired potential landscape.}

\begin{figure*}
    \centering
    \refstepcounter{figure}
    \label{fig:s2}
    \includegraphics[width=\linewidth]{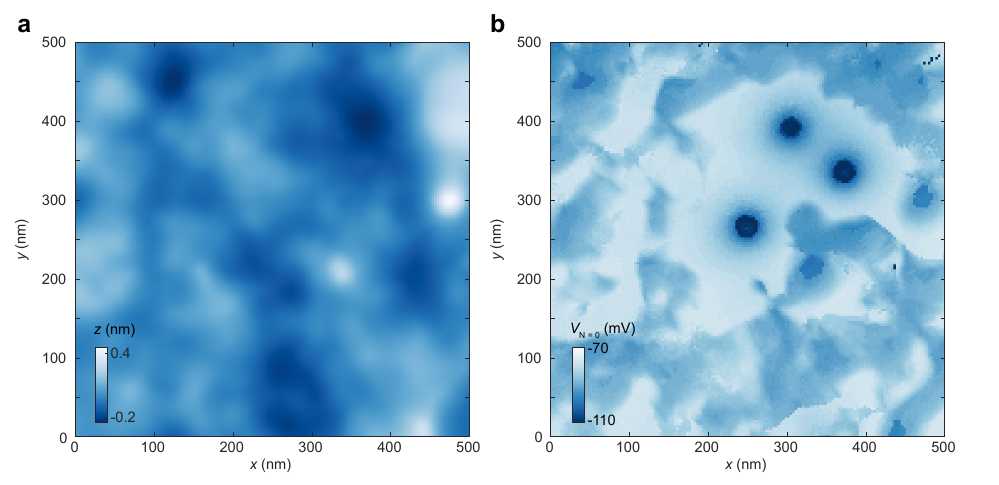}
\end{figure*}
\noindent\textbf{Fig. S2 \textbar{} Comparison between surface topography and simultaneously acquired potential landscape.} 
\textbf{a}, 500$\times$500 nm$^2$ topography on device A, where the surface is free of visible contamination and the maximum $z$ height variation is less than 1 nm. 
\textbf{b}, Real-space map of the ZLL peak bias voltage $V_{\mathrm{N = 0}}$ extracted from STS measurements at $B=12$ T, $\nu=2$, the shift of which reveals local electrostatic potential variations. The topography and STS data were obtained from the same measurement with a setpoint of $V_\text{B}=-250$ mV and $I_\text{t}=200$ pA, at $V_\text{G}=1.107$ V using AC oscillation of $V_\text{rms}=1$ mV. The topographic corrugations in \textbf{a} show little correlation with the potential variation in \textbf{b}.

\clearpage

\phantomsection
\addcontentsline{toc}{subsection}{Fig. S3 \textbar{} Visualizing valley symmetry-breaking of \textit{m} = 0 orbital on a surface defect.}

\begin{figure*}[!t]
    \centering
    \refstepcounter{figure}
    \label{fig:s3}
    \includegraphics[width=0.7\linewidth]{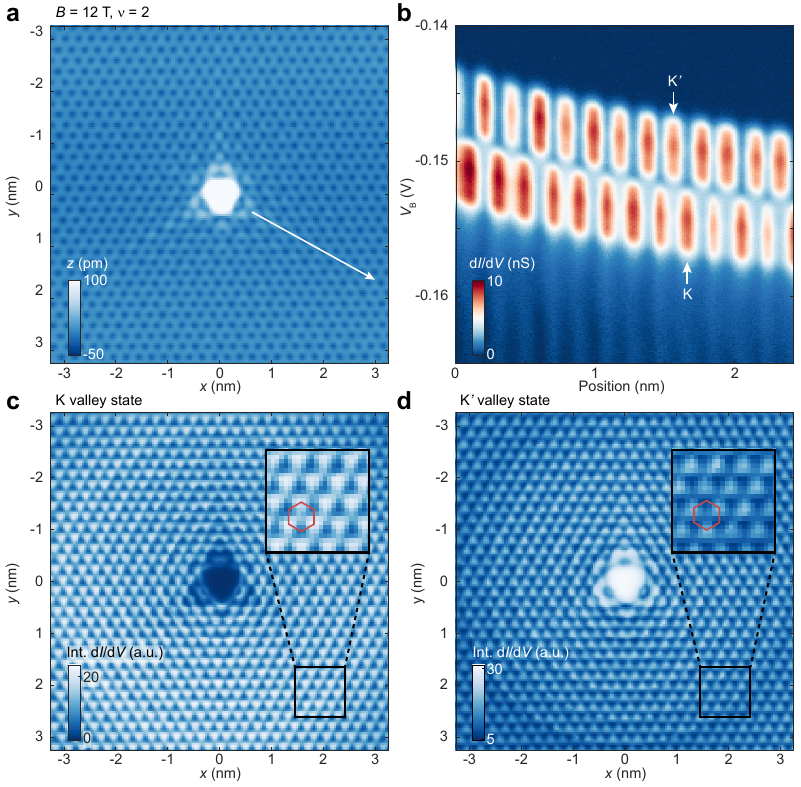}
\end{figure*}
\noindent\textbf{Fig. S3 \textbar{} Visualizing valley symmetry-breaking of \textit{m} = 0 orbital on a surface defect.}
\textbf{a}, Topography of a surface $+e$ defect in device A at $B=12$ T, $\nu=2$. The $\sqrt{3}\times\sqrt{3}R$30\textdegree\ pattern induced by intervalley scattering is evident around the defect center \cite{Pereira2008Modelingdisordergraphene, Ugeda2010MissingAtomSource}. 
\textbf{b}, High-resolution d$I$/d$V$ line cut through the surface $+e$ charged defect along the armchair direction of the graphene lattice, which is shown as a white arrow in \textbf{a}. This STS resolves the defect-induced valley splitting of $m = 0$ Landau orbits. Two branches of the ZLL peak corresponding to the $m=0$ state, denoted by $K$ and $K'$, are observed and exhibit a clear spatial offset of half a unit cell. Because valley and sublattice degrees of freedom are locked in the ZLL of monolayer graphene \cite{liuVisualizingBrokenSymmetry2022}, with the $K$ and $K'$ valleys residing on two sets of sublattices, respectively, the observed half-unit-cell shift between the two branches directly reveals their distinct valley character. The spin splitting within each valley branch is not clearly resolved in this data.
\textbf{c}, \textbf{d}, Integrated d$I$/d$V$ maps over $K$ (\textbf{c}) and $K'$ valley (\textbf{d}) states. Insets show zoomed-in results, revealing a clear sublattice polarization. The red hexagons in the insets mark a graphene honeycomb lattice. It is evident that the two valley states occupy different sublattice points.
Measurement condition: $V_\text{G} = 1.1$ V, $V_\text{B}=-200$ mV, $I_\text{t}=200$ pA, $V_\text{rms}=1$ mV.

\clearpage

\phantomsection
\addcontentsline{toc}{subsection}{Fig. S4 \textbar{} STS line cuts across surface and subsurface defects at various integer fillings.}

\begin{figure*}
    \centering
    \refstepcounter{figure}
    \label{fig:s4}
    \includegraphics[width=\linewidth]{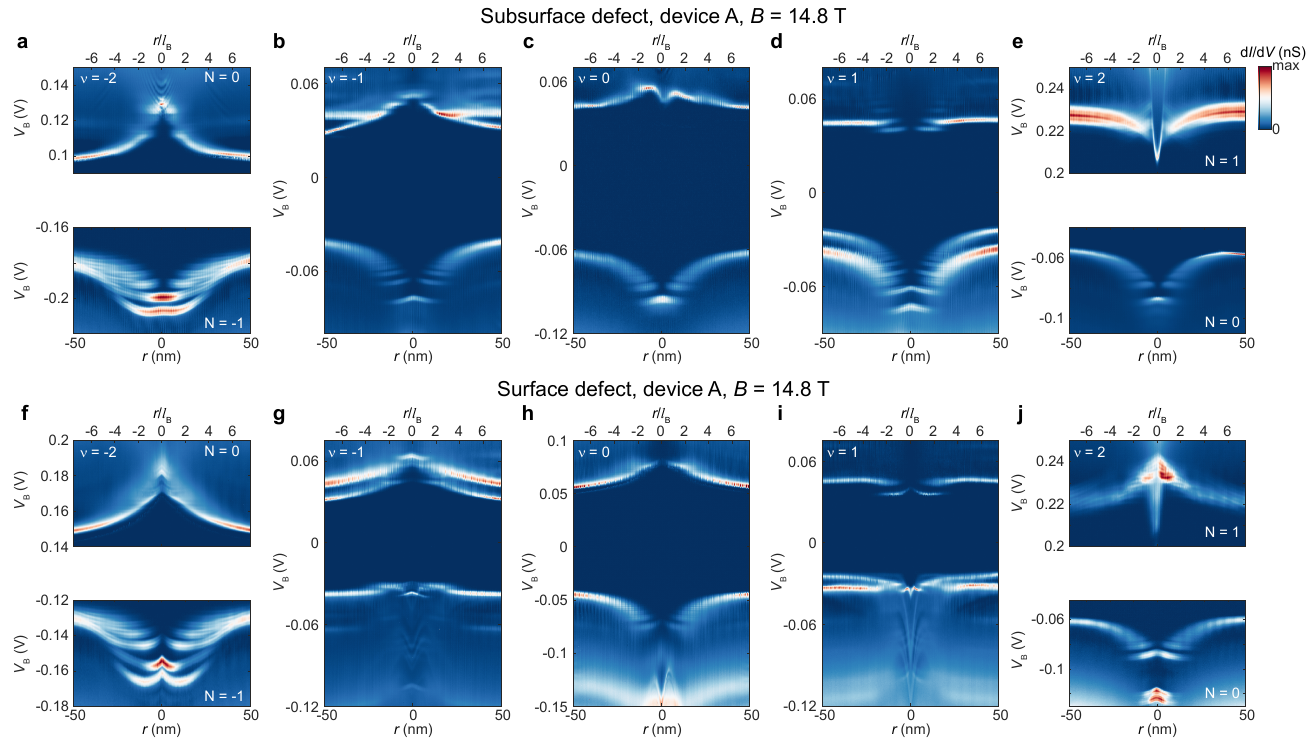}
\end{figure*}
\noindent\textbf{Fig. S4 \textbar{} STS line cuts across surface and subsurface defects at various integer fillings.}
\textbf{a}-\textbf{e}, STS line cuts across a subsurface defect in device A at $B=14.8$ T, $\nu=-2, -1, 0, 1, 2$.
\textbf{f}-\textbf{j}, STS line cuts across a surface defect in device A at $B=14.8$ T, $\nu=-2, -1, 0, 1, 2$. Note that in the bottom panels of \textbf{a} and \textbf{f}, the hole is tunneling into the N = -1 LL, and in the top panels of \textbf{e} and \textbf{j}, the electron is tunneling into the N = +1 LL.  In all other cases, tunneling occurs in the ZLL. To optimize signal-to-noise ratio and energy resolution, we use various measurement setpoints ranging from $V_\text{B}$ = -120 to -300 mV and $I_\text{t}$ = 0.2 to 2 nA with $V_\text{rms} = 0.5$ mV. Gate voltages used are the following: $V_\text{G}$ = -1.25 (\textbf{a}), -0.57 (\textbf{b}), -0.066 (\textbf{c}), 0.67 (\textbf{d}), 1.31 (\textbf{e}), -1.3 (\textbf{f}), -0.57 (\textbf{g}), 0.05 (\textbf{h}), 0.67 (\textbf{i}), and 1.32 (\textbf{j}) V. The electron-tunneling STS in \textbf{d} and \textbf{i} both exhibit flat LL peak energy profiles, at distances beyond two magnetic lengths, consistent with the defect being perfectly screened by a single bound electron. For the surface defect, the behavior in \textbf{i} is in stark contrast to that at $\nu=-1$ (\textbf{g}) and $\nu=0$ (\textbf{h}), where electron tunneling reveals clear signatures of overscreening. We attribute this difference to the fact that there is only one available empty state at $m = 0$ at $\nu=1$.  All hole excitation spectra, aside from \textbf{g} and \textbf{i}, are consistent with tunneling from an unscreened defect, while there appears to be a single electron bound in \textbf{g} and \textbf{i}.  See Fig. S\ref{fig:sim_v_-1} for theoretical modeling.
\clearpage

\phantomsection
\addcontentsline{toc}{subsection}{Fig. S5 \textbar{} Electrostatic simulation of the tip effect on LL spectra.}

\begin{figure*}
    \centering
    \refstepcounter{figure}
    \label{fig:tip_sim}
    \includegraphics[width=\linewidth]{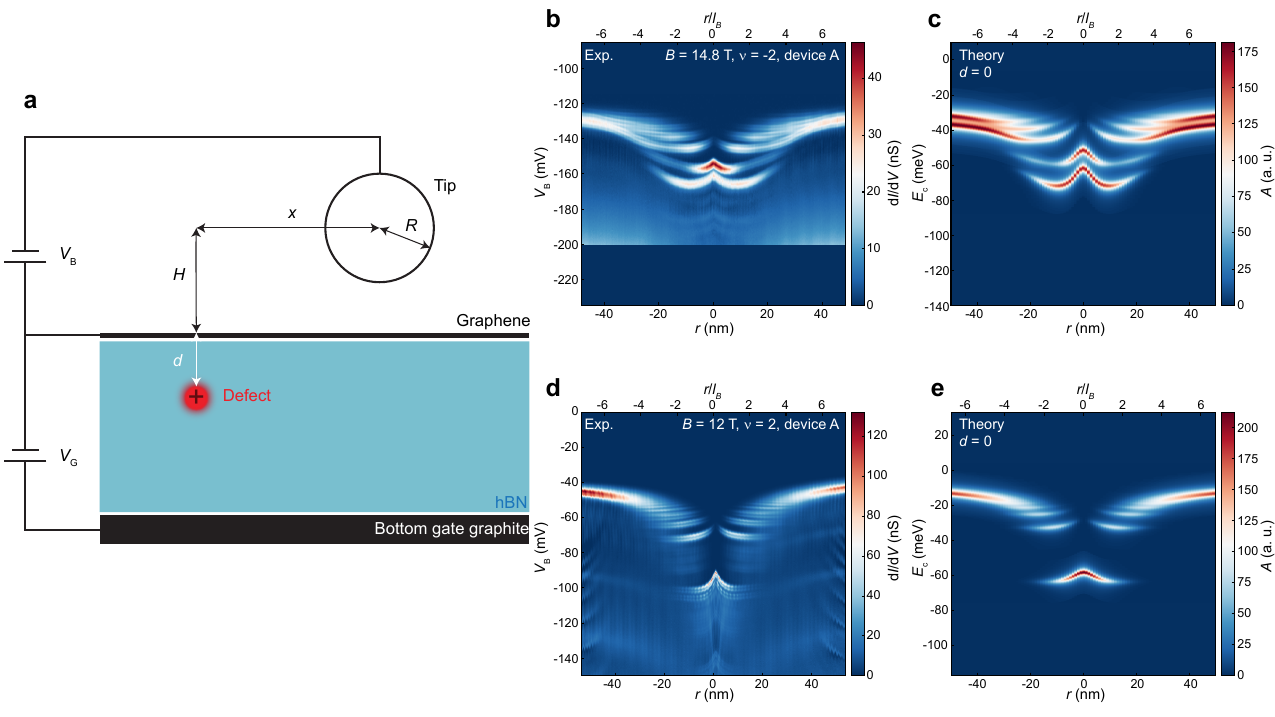}
\end{figure*}
\noindent\textbf{Fig. S5 \textbar{} Electrostatic simulation of the tip effect on LL spectra.}
\textbf{a}, Schematic illustration of our electrostatic model. The STM tip is modeled as a metal sphere of radius $R$, at height $H$ above the graphene layer and at horizontal distance $x$ from the defect center. The defect depth is $d$.
\textbf{b}, \textbf{c}, Comparison between experimental and calculated spectra of a surface $+e$ defect ($d=0$) at $B=14.8$ T, $\nu=-2$ in device A, featuring N = -1 LL. The cusp-like feature in the center is consistent with screening from the induced image charge in the tip sphere, while the upward bending away from the defect is caused by tip gating.
\textbf{d}, \textbf{e}, Comparison between experimental and calculated spectra of a surface $+e$ defect ($d=0$) at $B=12$ T, $\nu=2$ in device A, featuring ZLL in hole excitation. The tip screening is still prominent, but the upward bending due to tip gating is obscure. This can be attributed to the smaller applied tip bias voltage in \textbf{d} compared to \textbf{b} and the different spatial structure of N = -1 LL.
STM measurement conditions: (\textbf{b}) $V_\text{G}=-1.3$ V, $V_\text{B}=-200$ mV, $I_\text{t}=600$ pA, $V_\text{rms}=0.5$ mV; 
(\textbf{d}) $V_\text{G}=1.065$ V, $V_\text{B}=-200$ mV, $I_\text{t}=2$ nA, $V_\text{rms}=0.5$ mV.
Simulation parameters: (\textbf{c}) $H=4.5$ nm, $R=2$ nm, $d=0$, $V_t=-170$ mV;
(\textbf{e}) $H=5$ nm, $R=1.25$ nm, $d=0$, $V_t=-50$ mV.

\clearpage

\phantomsection
\addcontentsline{toc}{subsection}{Fig. S6 \textbar{} Spectral weight analysis across the surface +\textit{e} defect at v = -1 with different integrating windows.}

\begin{figure*}
    \centering
    \refstepcounter{figure}
    \label{fig:s6}
    \includegraphics[width=\linewidth]{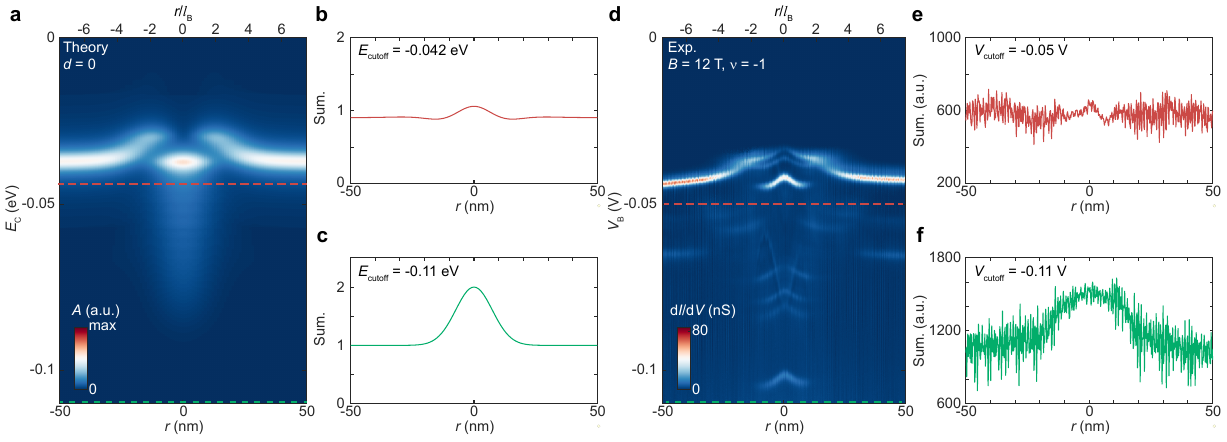}
\end{figure*}
\noindent\textbf{Fig. S6 \textbar{} Spectral weight analysis across the surface +\textit{e} defect at $\bm\nu$ = -1 with different integration windows.}
\textbf{a}, Spectral function simulation near a surface $+e$ defect ($d=0$) at $B=12$ T and $\nu=-1$, with one bound electron. The strong peak located at the defect center around $E_\text{c}=-45$ meV, corresponds to a process where the $m=0$ bound electron is removed. The branch that is flat away from the defect and tilts upward near the center involves the removal of an occupied $K\!\!\uparrow$ electron that feels the screened defect potential (Fig. 3a). The broad feature at $r=0$, extending roughly from $E_c=-100$ meV to $-45$ meV corresponds to the removal of a $K\!\!\uparrow$ electron at $m = 0$, which is coupled to a continuum of spin-wave states (see Fig. S\ref{fig:sim_v_-1} and theory Sec. \ref{subsubsec:theory_v-1}). The spectral weight of this continuum sums to one electron in the $m = 0$ orbital. In the calculation, a maximum angular momentum cutoff  ($m_\text{max}=420$ here) leads to dense but discrete energy levels rather than a continuous band.
\textbf{b}, \textbf{c}, Spectral weight integrated from \textbf{a} in the energy window between 0 and -42 meV (\textbf{b}) and between 0 and -110 meV (\textbf{c}). The cutoff energies are marked in \textbf{a} with corresponding colors. Excluding the continuum spectrum, the spectral weight is almost constant across the defect, at one electron per Landau orbital. Including the full spectrum, the spectral weight shows a pronounced peak and reaches two at the defect center, reflecting the extra bound $K\!\!\downarrow$ electron in addition to the uniformly occupied $K\!\!\uparrow$ band. 
\textbf{d}, Experimental STS line cut across a surface $+e$ defect in device A at $B=12$ T and $\nu=-1$.
\textbf{e}, \textbf{f}, Experimental spectral weight with energy integration  windows nearly identical to those in \textbf{b} and \textbf{c}. The cutoff bias voltages are similarly marked in \textbf{d}. The general behavior is similar between theory and experiment. However, due to the non-zero d$I$/d$V$ background in the experiment, the spectral weight profile in \textbf{f} is not identical to the theory in \textbf{c}. This non-zero background may be attributed to bias-dependent tunneling matrix and extra spectral peaks of an undetermined origin.
STM measurement conditions: $V_\text{G}=-0.45$ V, $V_\text{B}=-200$ mV, $I_\text{t}=2$ nA, $V_\text{rms}=0.5$ mV.
\clearpage

\phantomsection
\addcontentsline{toc}{subsection}{Fig. S7 \textbar{} Simulated spectra for surface and subsurface +\textit{e} defect at v = -1.}

\begin{figure*}[!t]
    \centering
    \refstepcounter{figure}
    \label{fig:sim_v_-1}
    \includegraphics[width=\linewidth]{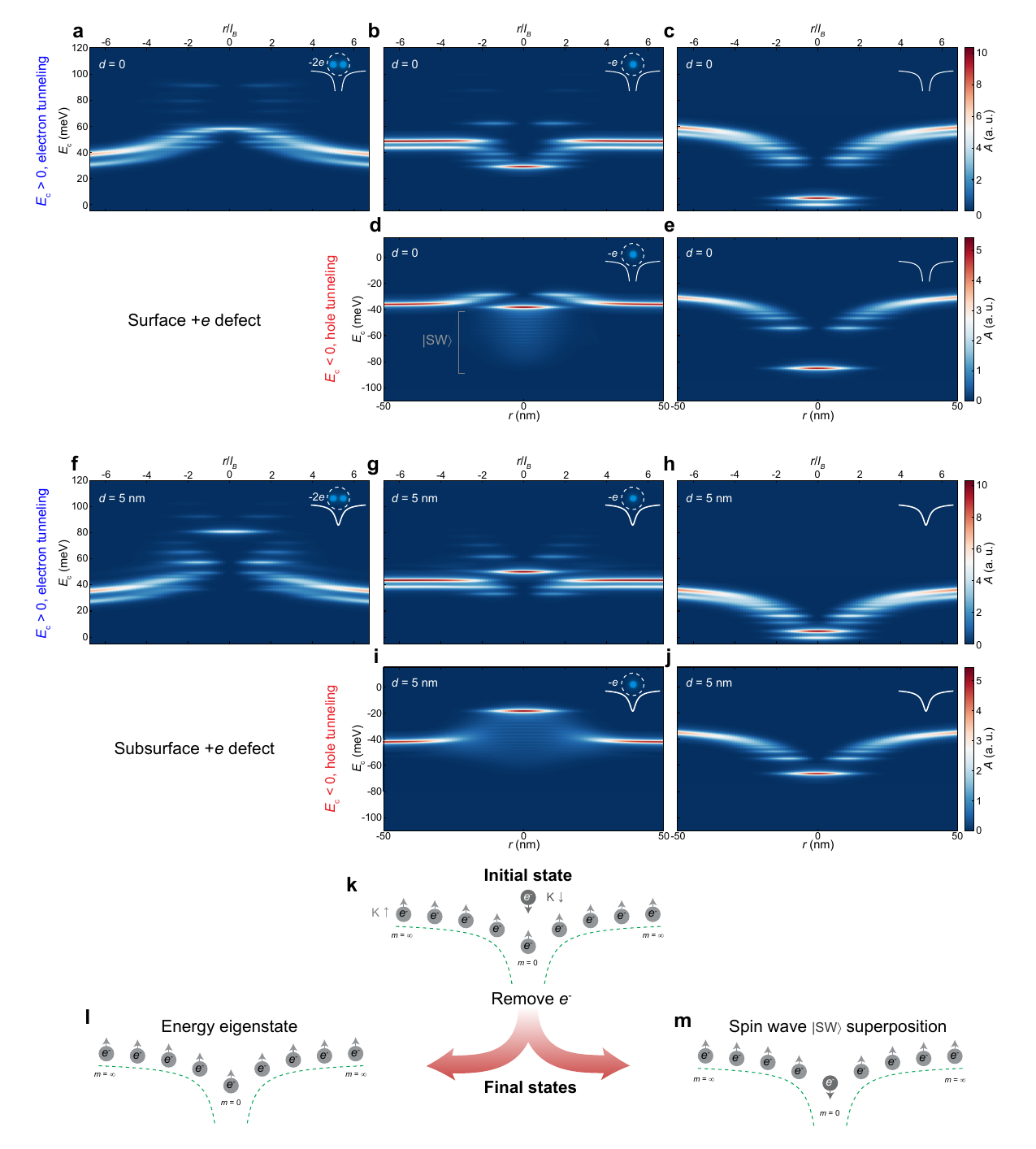}
\end{figure*}
\noindent\textbf{Fig. S7 \textbar{} Simulated spectra for surface and subsurface +\textit{e} defects at $\bm \nu$ = -1.}
\textbf{a}-\textbf{e}, Simulated electron ($E_\mathrm{c}>0$) and hole ($E_\mathrm{c}<0$) tunneling spectra across a surface $+e$ defect ($d=0$) with two (\textbf{a}), one (\textbf{b} and \textbf{d}), or zero (\textbf{c} and \textbf{e}) bound electrons at $B=12$ T. Top row (\textbf{a} to \textbf{c}) is electron tunneling and bottom row (\textbf{d} and \textbf{e}) is hole tunneling.
\textbf{f}-\textbf{j}, Simulated electron and hole tunneling spectra across a subsurface $+e$ defect ($d=5$ nm) at $B = 12$ T, similar to \textbf{a} to \textbf{e}.
Insets illustrate electron binding number and defect Coulomb potential (white curve). Calculation is performed following the theoretical model in Sec. \ref{subsubsec:theory_v-1}.

The empty states of $\nu=-1$ consist of three isospin flavors, namely $K\!\!\downarrow$, $K'\!\!\uparrow$, and $K'\!\!\downarrow$. $K'\!\!\uparrow$ and $K'\!\!\downarrow$ are near degenerate and are split from $K\!\!\downarrow$ by a valley anisotropy gap. In our calculation, we included a flavor($f$)-specific energy term $\Delta E_f$ to account for the valley splitting caused by anisotropic interactions and the substrate.  

Electron binding produces distinct signatures in the simulated spectra. The number of bound electrons can first be inferred from the slope of the spectral peak several magnetic lengths away from the defect, which can be interpreted as the potential landscape for an added electron: two bound electrons ($-2e$) overscreen the defect potential and produce an energy hill with a peak at the defect center, one bound electron ($-e$) nearly perfectly screens the defect potential and yields an almost flat energy, whereas the absence of bound electrons preserves the original $+e$ defect potential, with a distinct funnel shape, centered at the defect location. When a single electron is bound, it occupies the $K\!\!\downarrow$ state with $m=0$. In contrast, when two electrons are bound, the second electron may occupy the $K'\!\!\uparrow$ or $K'\!\!\downarrow$ state, or even form a superposition state. In our model, we consider only one specific flavor configuration. However, it is plausible that the second bound quasiparticle forms a skyrmion rather than an $m=0$ electron, which may account for the discrepancy with the experimental results (Fig. 3c).

\textbf{k}-\textbf{m}, Illustration of the hole-tunneling processes for a one-electron bound state. The initial state consists of a bound $K\!\!\downarrow$, $m=0$ electron in addition to a fully occupied $K\!\!\uparrow$ band. Upon hole tunneling at $r = 0$, the STM tip may remove the bound $K\!\!\downarrow$ electron, returning the system to the energy eigenstate shown in \textbf{l} and producing a sharp peak at the defect center (see \textbf{d} and \textbf{i}). Alternatively, the tunneling may remove a $K\!\!\uparrow$, $m=0$ electron, producing a state with uniform charge density and a flipped spin at the center (\textbf{m}). This resulting state is not an eigenstate of the Hamiltonian, but rather a superposition of many spin wave (magnon) excitations. In the thermodynamic limit, these excitations form a continuum; however, they appear as dense but discrete states in \textbf{d} due to the finite angular momentum basis used in the calculation. The green dashed curves represent the defect Coulomb potential.

Simulation parameters: (\textbf{a} and \textbf{f}) $d=0/5$ nm, $\Delta E_{K\uparrow}=4$ meV, $\Delta E_{K\downarrow}=2$ meV, $\Delta E_{K'\uparrow}=10$ meV, $E_{K'\downarrow}=10$ meV; 
(\textbf{b} and \textbf{g}) $d=0/5$ nm, $\Delta E_{K\uparrow}=0$ meV, $\Delta E_{K\downarrow}=6$ meV, $\Delta E_{K'\uparrow}=11$ meV, $E_{K'\downarrow}=11$ meV;
(\textbf{c} and \textbf{h}) $d=0/5$ nm, $\Delta E_{K\uparrow}=0$ meV, $\Delta E_{K\downarrow}=6$ meV, $\Delta E_{K'\uparrow}=11$ meV, $E_{K'\downarrow}=11$ meV;
(\textbf{d} and \textbf{i}) $d=0/5$ nm, angular momentum cutoff $m_\text{cutoff}= 420$, $\Delta E_{K\uparrow}=4$ meV, $\Delta E_{K\downarrow}=2$ meV, $\Delta E_{K'\uparrow}=10$ meV, $E_{K'\downarrow}=10$ meV; 
(\textbf{e} and \textbf{j}) $d=0/5$ nm, $\Delta E_{K\uparrow}=0$ meV, $\Delta E_{K\downarrow}=6$ meV, $\Delta E_{K'\uparrow}=11$ meV, $E_{K'\downarrow}=11$ meV.
\clearpage

\phantomsection
\addcontentsline{toc}{subsection}{Fig. S8 \textbar{} Gate-dependence of STS line cuts spanning the v = -1 quantum Hall ferromagnetic states for subsurface and surface defects at \textit{B} = 14.8 T in device A.}

\begin{figure*}
    \centering
    \refstepcounter{figure}
    \label{fig:s8}
    \includegraphics[width=\linewidth]{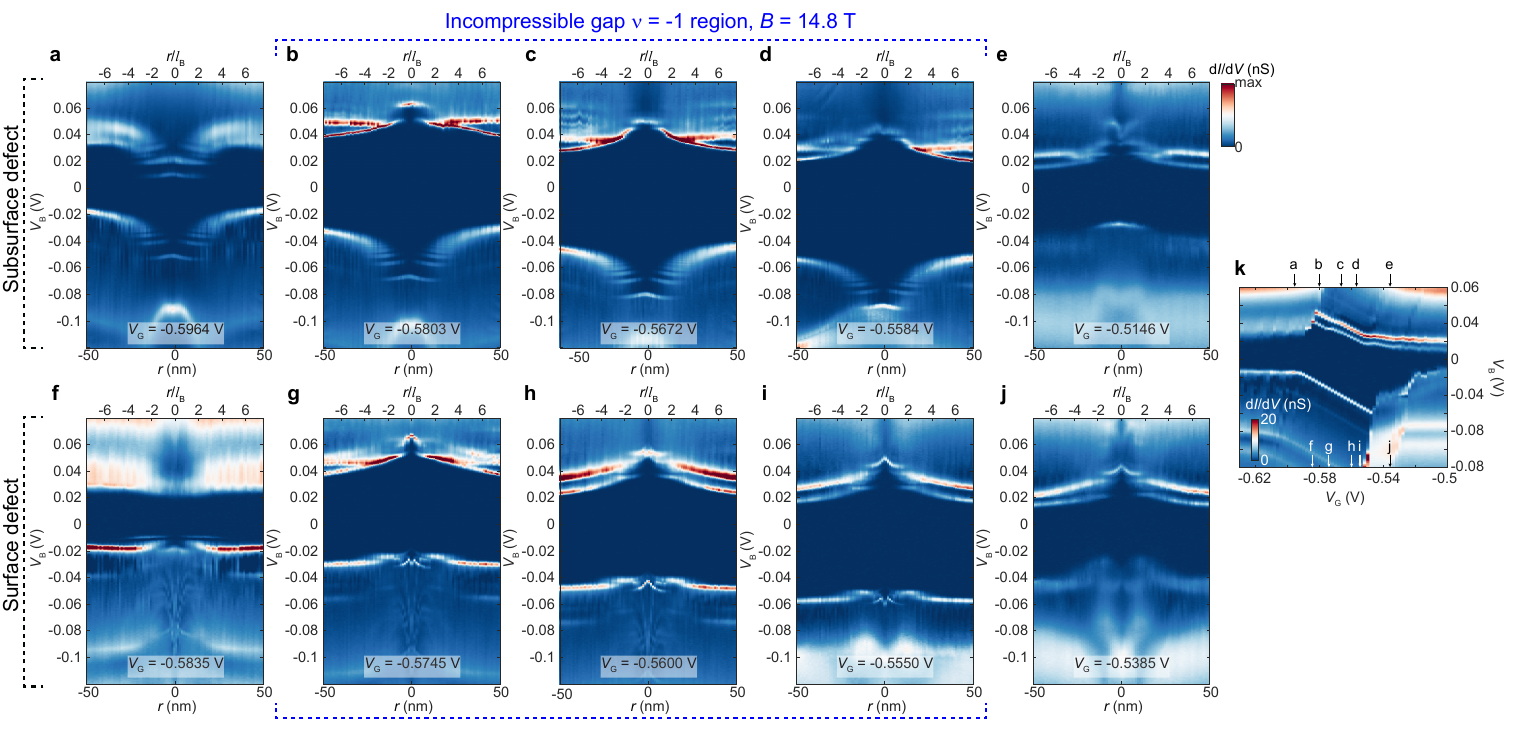}
\end{figure*}
\noindent\textbf{Fig. S8 \textbar{} Gate-dependence of STS line cuts spanning the $\bm\nu$ = -1 quantum Hall ferromagnetic states for subsurface and surface defects at \textit{B} = 14.8 T in device A.}
\textbf{a}-\textbf{e}, STS measured across a subsurface $+e$ defect at various gate voltages spanning the $\nu=-1$ state. \textbf{a} and \textbf{e} correspond to gate voltages outside the incompressible state, whereas panels \textbf{b} to \textbf{d} are acquired within the incompressible regime (see \textbf{k}). The tip height was set using $V_\text{B}=-120$ mV and $I_\text{t}=600$ pA, after which the feedback was turned off for STS acquisition.
\textbf{f}-\textbf{j}, Surface $+e$ defect STS across compressible (\textbf{f} and \textbf{j}) and incompressible states (\textbf{g} to \textbf{i}). Tip heights are set by turning off feedback loop at $V_\text{B}$ = -170 mV, $I_\text{t}$ = 1 nA. 
\textbf{k}, Gate-dependent point spectrum across the $\nu=-1$ quantum Hall ferromagnetic state taken away from strong defect potentials. The diamond-shaped region corresponds to the incompressible regime. Within the incompressible gap, electron and hole tunneling pin the local chemical potential to the top and bottom of the $\nu=-1$ gap, which enlarges the apparent spectral gap (see Fig. S\ref{fig:pinning} for detailed discussion). The gate voltages corresponding to panels \textbf{a} to \textbf{j} are indicated. For both surface and subsurface defects within the incompressible regime (\textbf{b} to \textbf{d} and \textbf{g} to \textbf{i}, respectively), the STS spectrum shifts rigidly with gate voltage $V_\text{G}$ without significant changes in the spectral features, indicating that the bound electron number does not change with gate voltages within the incompressible state.
\clearpage

\phantomsection
\addcontentsline{toc}{subsection}{Fig. S9 \textbar{} Tunneling-current-induced chemical potential pinning and its effect on the apparent quantum Hall gap in STS.}

\begin{figure*}
    \centering
    \refstepcounter{figure}
    \label{fig:pinning}
    \includegraphics[width=0.7\linewidth]{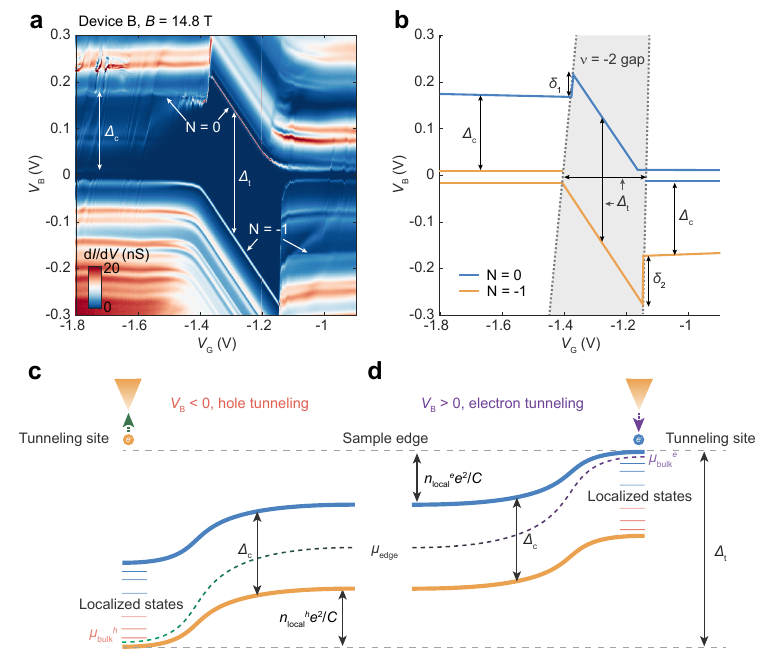}
\end{figure*}
\noindent\textbf{Fig. S9 \textbar{} Tunneling-current-induced chemical potential pinning and its effect on the apparent quantum Hall gap in STS.}
\textbf{a}, \textbf{b}, Gate-dependent point spectrum around the $\nu=-2$ quantum Hall state measured in device B at $B=14.8$ T away from strong defect potentials. The energy gap between the N = 0 and N = -1 LLs observed in the compressible regime is $\Delta_\text{c}\approx170$ meV, consistent with the theoretical cyclotron gap. In contrast, the apparent tunneling gap between the two LLs is enlarged to $\Delta_\text{t}\approx290$ meV within the incompressible $\nu = -2$ quantum Hall state. A schematic of the corresponding LL energy evolution is shown in \textbf{b}. The spectrum was acquired using the constant-height-setpoint approach described in Methods, with a setpoint bias of $V_\text{B}=-1$ V and a lock-in excitation amplitude of $V_\text{rms}=1$ mV.

The origin of the enhanced energy separation between LLs immediately below and above the Fermi energy in incompressible states has remained a mystery since these features were first observed in STM experiments \cite{liuVisualizingBrokenSymmetry2022, Hu2025Highresolutiontunnellingspectroscopy}. Here, we argue that this enhancement arises naturally from tunneling-current-induced chemical potential pinning. In the Main Text, we discussed how the local chemical potential becomes pinned when tunneling into an insulating state. Specifically, during electron tunneling, the local chemical potential near the tunneling site is pinned to the upper edge of the gap, independent of gate voltages. This is because injected electrons become trapped in localized in-gap states until their energy reaches an extended state connected to the sample edge. Conversely, for hole tunneling, the in-gap states are emptied and the local chemical potential is pinned to the lower edge of the gap.

\textbf{c}, \textbf{d} illustrate the resulting LL energy shifts caused by the accumulation of localized charges during the tunneling process. In hole tunneling (\textbf{c}), LLs near the tunneling site are shifted downward in energy relative to the LL energies at the sample edge by an amount $n_{\text{local}}^{h}e/C$, where $n_{\text{local}}^{h}$ is the excess hole density arising from tunneling-caused local chemical potential pinning. Physically, $n_{\text{local}}^{h}$ corresponds to the density of localized in-gap states (orange lines) between $\mu_{\text{edge}}$ and the mobility edge of the lower LL, and $C$ is the capacitance between the sample and the bottom-gate. Similarly, during electron tunneling (\textbf{d}), the accumulation of localized electrons shifts the LLs upward by $n_{\text{local}}^{e}e/C$, where $n_{\text{local}}^{e}$ is the density of excess electrons induced by tunneling. This rise of LL energy can also be understood as the Coulomb repulsion between incoming electrons from the tip and the extra localized charge of pinned in-gap states. Consequently, the apparent tunneling gap across the Fermi energy becomes $\Delta_\text{t}=\Delta_\text{c} +n_{\text{local}}e/C$
, where $n_{\text{local}}=n_{\text{local}}^{e}+n_{\text{local}}^{h}$ is the total density of localized in-gap states. Therefore, the enhancement of the apparent tunneling gap beyond the cyclotron gap is directly related to the density of in-gap states.

This interpretation explains several features observed in \textbf{a}. First, within the incompressible state, the LL peaks shift in energy with changing gate voltages with a one-to-one slope. Without considering the chemical-potential pinning effect, this behavior would be surprising, as it would imply the absence of in-gap states while the chemical potential traverses the gap. Considering the chemical-potential pinning effect, however, such behavior arises naturally even when localized in-gap states are present, because the occupation of these states, and hence the local charge density, remains independent of gate voltage. Consequently, within the incompressible regime marked by the gray region in \textbf{b}, varying the gate voltage shifts the electrostatic potential of graphene without changing its charge density, making the system behave as if no in-gap states were present.

As the system enters the compressible regime, the local chemical potential pinning effect disappears because the sample is no longer insulating. During electron tunneling, the local chemical potential is pinned to the upper edge of the gap within the incompressible state, allowing the corresponding LL energies to connect smoothly to those on the $\nu>-2$ side. However, when the gate voltage is lowered such that $\nu<-2$, the chemical potential abruptly jumps from the upper to the lower edge of the cyclotron gap, producing a discontinuity of magnitude $\delta_1$. Conversely, during hole tunneling, an analogous discontinuity with magnitude $\delta_2$ appears on the higher filling side of the incompressible state. In the ideal case, one expects $\delta_1=\delta_2={n_{\text{local}}e}/{C} $. Experimentally, however, a small asymmetry is observed between $\delta_1$ and $\delta_2$, which may arise from tip-gating effects or many-body correlation effects.

\clearpage

\phantomsection
\addcontentsline{toc}{subsection}{Fig. S10 \textbar{} DMRG simulations of electron-excitation STS across -/+e defect at \textit{B} = 14.8 T for the hole filling v* = 1/3 fractional quantum Hall state.}

\begin{figure*}
    \centering
    \refstepcounter{figure}
    \label{fig:fractional_sim}
    \includegraphics[width=\linewidth]{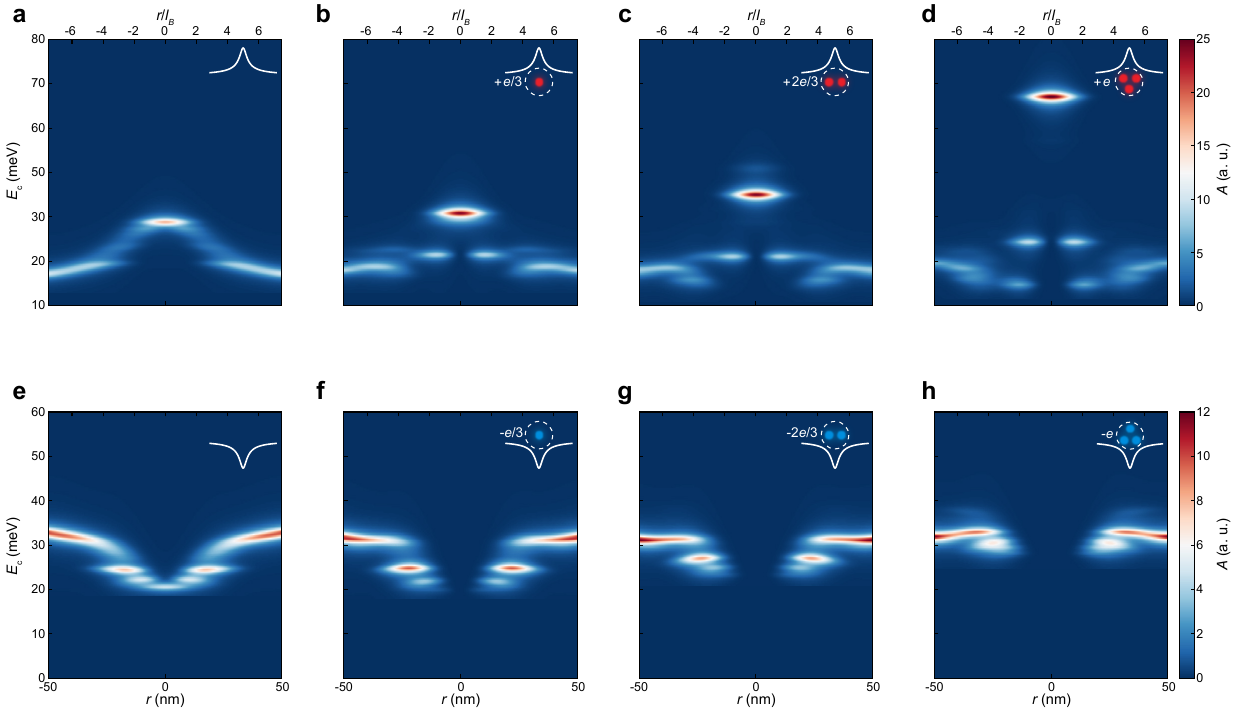}
\end{figure*}
\noindent\textbf{Fig. S10 \textbar{} DMRG simulations of electron-excitation STS across $\bm{\mp e}$ defect at \textit{B} = 14.8 T for the hole filling  $\bm{\nu^*=1/3}$ fractional quantum Hall state.} 
\textbf{a}-\textbf{d}, Simulated electron-excitation spectra across a $-e$ defect at various depths, corresponding to 0 (\textbf{a}), 1 (\textbf{b}), 2 (\textbf{c}), and 3 (\textbf{d}) bound $+e/3$ anyon quasi-holes. 
\textbf{e}-\textbf{h}, Same as \textbf{a} to \textbf{d}, but for a $+e$ defect that binds 0 (\textbf{e}), 1 (\textbf{f}), 2 (\textbf{g}), 3 (\textbf{h}) anyon quasi-electrons. Insets show schematic defect Coulomb potentials (white curves) and the inferred numbers of bound quasiparticles. Red dots denote $+e/3$ quasi-holes, while blue dots denote $-e/3$ quasi-electrons.

The simulations are intended to model the $\nu=5/3$ electron-tunneling data shown in Fig. 4c and e, which correspond to a hole filling of $\nu^*=1/3$ considered here. Owing to the particle-hole symmetry of the model, reversing the signs of the defect charge ($\pm e$), filling factor ($\nu = -5/3$), and energy (flip $E_\text{c}$ axis and consider hole excitation STS) produces identical results. All DMRG calculations presented in this figure use models with $N=6$ holes, whereas the calculations shown in Fig. 4b and d employ a larger model with $N=8$ holes.

We track the evolution of the spectra as a function of defect depth $d$ and the number of manually introduced quasi-holes $N_\text{qh}$ (negative $N_\text{qh}$ refers to number of quasi-electrons) relative to the $\nu=1/3$ Laughlin state. In the Haldane-sphere geometry, this is achieved by varying the magnetic monopole flux number (see references \cite{Arciniaga2016, Haldane1985} and theory Sec. \ref{subsubsec:theory_v1/3}). The resulting spectra fall into distinct categories corresponding to different anyon binding numbers, which we summarize and present here.

Simulation parameters: (\textbf{a}) $N_\text{qh}=0$, $d=14.61$ nm;
(\textbf{b}) $N_\text{qh}=1$, $d=7.41$ nm;
(\textbf{c}) $N_\text{qh}=2$, $d=4.42$ nm;
(\textbf{d}) $N_\text{qh}=3$, $d=0$ nm;
(\textbf{e}) $N_\text{qh}=0$, $d=12.78$ nm;
(\textbf{f}) $N_\text{qh}=-1$, $d=3.77$ nm;
(\textbf{g}) $N_\text{qh}=-2$, $d=0$ nm;
(\textbf{h}) $N_\text{qh}=-3$, $d=0$ nm;
(Fig. 4b), $N_\text{qh}=3$, $d=3$ nm;
(Fig. 4d), $N_\text{qh}=-3$, $d=0.92$ nm.

\clearpage

\phantomsection
\addcontentsline{toc}{subsection}{Fig. S11 \textbar{} Comparison between STS across a +\textit{e} charged defect at v = -5/3 and a -\textit{e} charged defect at v = 5/3, showing particle-hole symmetry.}

\begin{figure*}
    \centering
    \refstepcounter{figure}
    \label{fig:negative_inverted}
    \includegraphics[width=\linewidth]{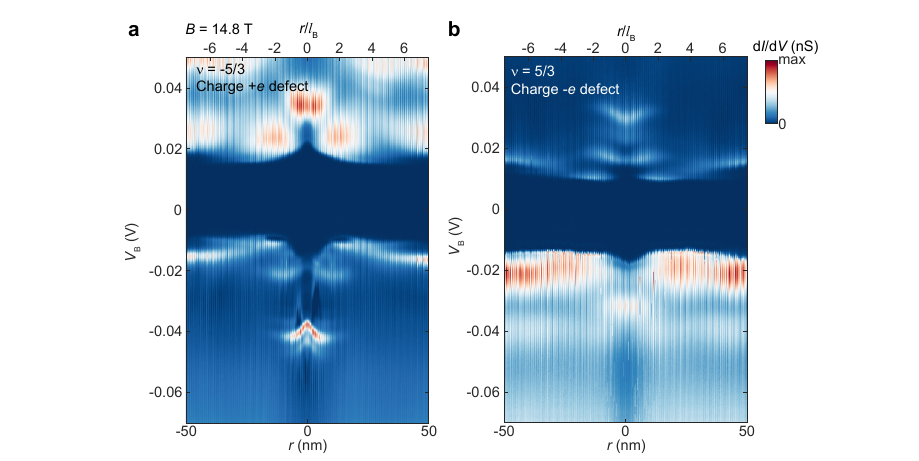}
\end{figure*}
\noindent\textbf{Fig. S11 \textbar{} Comparison between STS across a +\textit{e} charged defect at $\bm \nu$ = -5/3 and a -\textit{e} charged defect at $\bm \nu$ = 5/3, showing particle-hole symmetry.}
\textbf{a}, \textbf{b}, STS across a $+e$ defect of depth $d \approx 1.04$ nm at $\nu=-5/3$  (\textbf{a}) and a $-e$ defect of depth $d \approx 3$ nm at $\nu=5/3$ (\textbf{b}); both were taken in device B at $B = 14.8$ T. One can clearly note the symmetry of these two measurements: $\frac{\mathrm{d}I}{\mathrm{d}V}(V_\text{B}; \nu = 5/3, -e) = \frac{\mathrm{d}I}{\mathrm{d}V}(-V_\text{B}; \nu = -5/3, +e)$, which follows the expected particle-hole symmetry. The slight difference between the two after mirroring may be attributed to the difference in the defect depth.
Measurement conditions: (\textbf{a}) Tip height setpoint: $V_\text{G}=1.57$ V, $V_\text{B}=-400$ mV, $I_\text{t}=100$ pA. Tip is then lowered by $\Delta z = -260$ pm for STS measurement at $V_\text{G}=-1.639$ V with $V_\text{rms}=0.5$ mV.
(\textbf{b}) $V_\text{G}=1.739$ V, $V_\text{B}=-1$ V, $I_\text{t}=100$ pA, $\Delta z=-370$ pm, $V_\text{rms}=0.5$ mV.

\clearpage

\phantomsection
\addcontentsline{toc}{subsection}{Fig. S12 \textbar{} Spectral weight extraction for the spectral weight plots shown in Fig. 3 and 4.}

\begin{figure*}
    \centering
    \refstepcounter{figure}
    \label{fig:s12}
    \includegraphics[width=\linewidth]{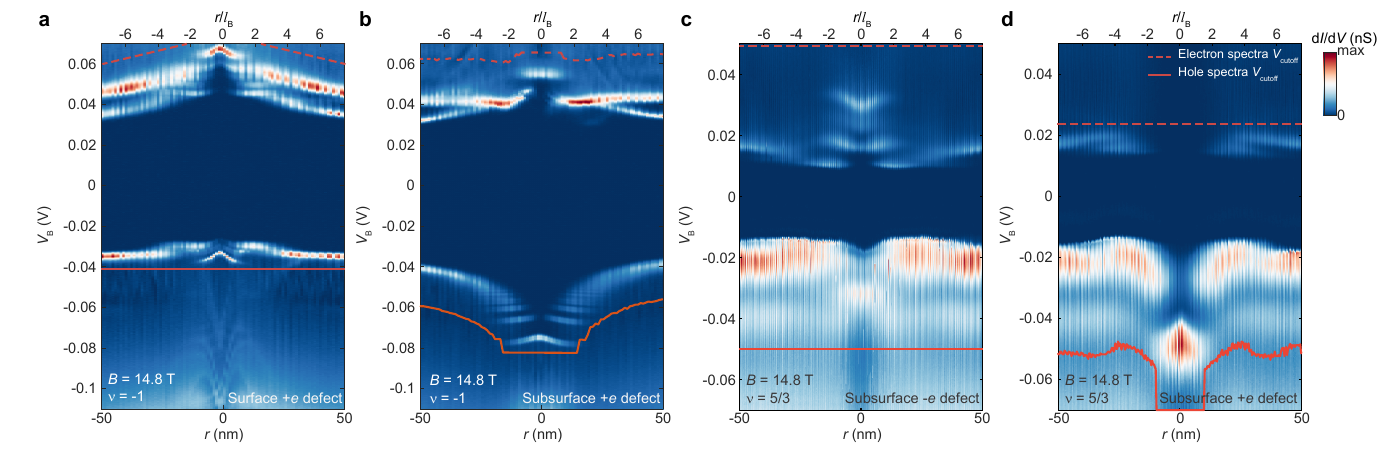}
\end{figure*}
\noindent\textbf{Fig. S12 \textbar{} Spectral weight extraction for the spectral weight plots shown in Fig. 3 and 4.} 
\textbf{a}, \textbf{b}, Raw STS data acquired at $B=14.8$ T and $\nu=-1$ across a surface $+e$ defect (\textbf{a}) and a subsurface $+e$ defect (\textbf{b}). Spectral weight extracted from these data is shown in Fig. 3b and 3d, respectively.
\textbf{c}, \textbf{d}, Raw STS data acquired at $B=14.8$ T and $\nu=5/3$ across a subsurface $-e$ defect (\textbf{c}) and a subsurface $+e$ defect (\textbf{d}), reproduced from Fig. 4c and e. These data are used to obtain the spectral-weight profiles shown in Fig. 4f.

For electron excitations, the spectral weight is obtained by integrating the d$I$/d$V$ signal from zero energy to the upper cutoff marked by the red dashed lines. For hole excitations, the spectral weight is obtained by integrating from the lower cutoff indicated by the solid red lines to zero energy.

Measurement conditions: (\textbf{a}) $V_\text{G}=-0.57$ V, $\Delta z=-330$ pm, $V_\text{rms}=0.5$ mV, reference setpoint $V_\text{G}=-0.54$ V, $V_\text{B}=1$ V, $I_\text{t}=100$ pA; 
(\textbf{b}) $V_\text{G}=-0.57$ V, $\Delta z=-330$ pm, $V_\text{rms}=0.5$ mV, reference setpoint $V_\text{G}=-0.54$ V, $V_\text{B}=1$ V, $I_\text{t}=100$ pA; 
(\textbf{c}) $V_\text{G}=1.739$ V, $\Delta z=-370$ pm, $V_\text{rms}=0.5$ mV, reference setpoint $V_\text{G}=1$ V, $V_\text{B}=-1$ V, $I_\text{t}=100$ pA; 
(\textbf{d}) $V_\text{G}=1.739$ V, $V_\text{B}=-1$ V, $I_\text{t}=100$ pA, $\Delta z=-370$ pm, $V_\text{rms}=0.5$ mV.

\clearpage

\phantomsection
\addcontentsline{toc}{subsection}{Fig. S13 \textbar{} Characterization of the switchable subsurface -\textit{e} defect in Fig. 4.}

\begin{figure*}
    \centering
    \refstepcounter{figure}
    \label{fig:gate_switch_defect}
    \includegraphics[width=\linewidth]{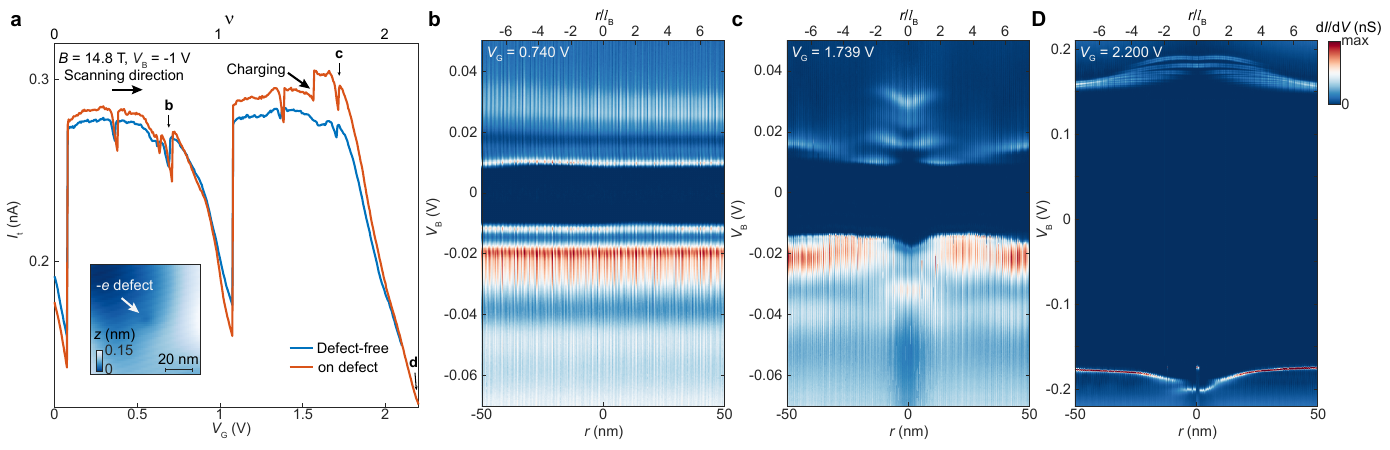}
\end{figure*}
\noindent\textbf{Fig. S13 \textbar{} Characterization of the switchable subsurface $-e$ defect shown in Fig. 4.} 
\textbf{a}, Tunneling current measured at $V_\text{B}=-1$ V as a function of gate voltage directly above the subsurface defect (orange curve) and far away from the defect (blue curve). Both measurements were performed on device B at a fixed tip height. For clarity, the off-defect current trace is vertically offset by 50 pA. The current dips correspond to integer and fractional incompressible states (for example, $\nu=0$ near $V_\text{G}=0$ V and $\nu=1$ near $V_\text{G}=1$ V). The on-defect current exhibits an abrupt jump at $V_\text{G}^{}\approx1.56$ V, corresponding to a sudden change in the defect charge state. For $V_\text{G}>V_\text{G}^*$, the defect carries a charge of $-e$, whereas it is charge neutral for $V_\text{G}<V_\text{G}^{*}$. Inset: topographic image of the charged defect. From the orbital splitting pattern, we estimate the defect depth to be $d\approx3$ nm (see Fig. S\ref{fig:defect_split}).

\textbf{b}-\textbf{d}, STS line cuts across the defect acquired at different gate voltages spanning the charging transition. At $V_\text{G}=0.74$ V (\textbf{b}), the tunneling spectrum shows little spatial variation and no discernible signature of the defect, indicating that the defect is neutral at this gate voltage. In contrast, for $V_\text{G}>V_\text{G}^{*}$ (\textbf{c} and \textbf{d}), the influence of the impurity potential becomes clearly visible. \textbf{d} shows the spectrum at $\nu=2$, exhibiting an orbital splitting pattern that is inverted from that observed for a positively charged $+e$ defect at $\nu=-2$ (e.g., Fig. S\ref{fig:tip_sim}b). This behavior further confirms that the defect carries a charge of $-e$ (see also Fig. S\ref{fig:negative_inverted} and S\ref{fig:defect_split}).

This switchable defect provides a unique opportunity to measure the spectral weight with the highest accuracy. In conventional STS measurements, the extracted spectral weight may be affected by the setpoint effect, whereby the tip-sample distance varies in response to changes in the tunneling current induced, for example, by a charged defect. Although we minimize this effect by setting the tip height in a compressible state where the defect potential is screened, residual systematic errors associated with the setpoint effect may still remain. For the spectral-weight data shown in Fig. 4f for the $-e$ defect, the tip-height trace was acquired at a gate voltage $V_\text{G}<V_\text{G}^{*}$, where the defect is charge neutral. As a result, the tip height is not affected by the charge neutral defect, providing a particularly reliable measurement of spectral weight free from setpoint-induced artifacts.
 
Measurement conditions: (\textbf{b}) $V_\text{G}=0.74$ V, $V_\text{B}=-1$ V, $I_\text{t}=100$ pA, $\Delta z=-370$ pm; 
(\textbf{c}) $V_\text{G}=1.739$ V, $\Delta z=-370$ pm, $V_\text{rms}=0.5$ mV, reference setpoint $V_\text{G}=1$ V, $V_\text{B}=-1$ V, $I_\text{t}=100$ pA;
(\textbf{d}) $V_\text{G}=2.2$ V, $V_\text{B}=-1$ V, $I_\text{t}=100$ pA, $\Delta z=-275$ pm. 

\clearpage

\phantomsection
\addcontentsline{toc}{subsection}{Fig. S14 \textbar{} Estimation of subsurface defect depths.}

\begin{figure*}
    \centering
    \refstepcounter{figure}
    \label{fig:defect_split}
    \includegraphics[width=\linewidth]{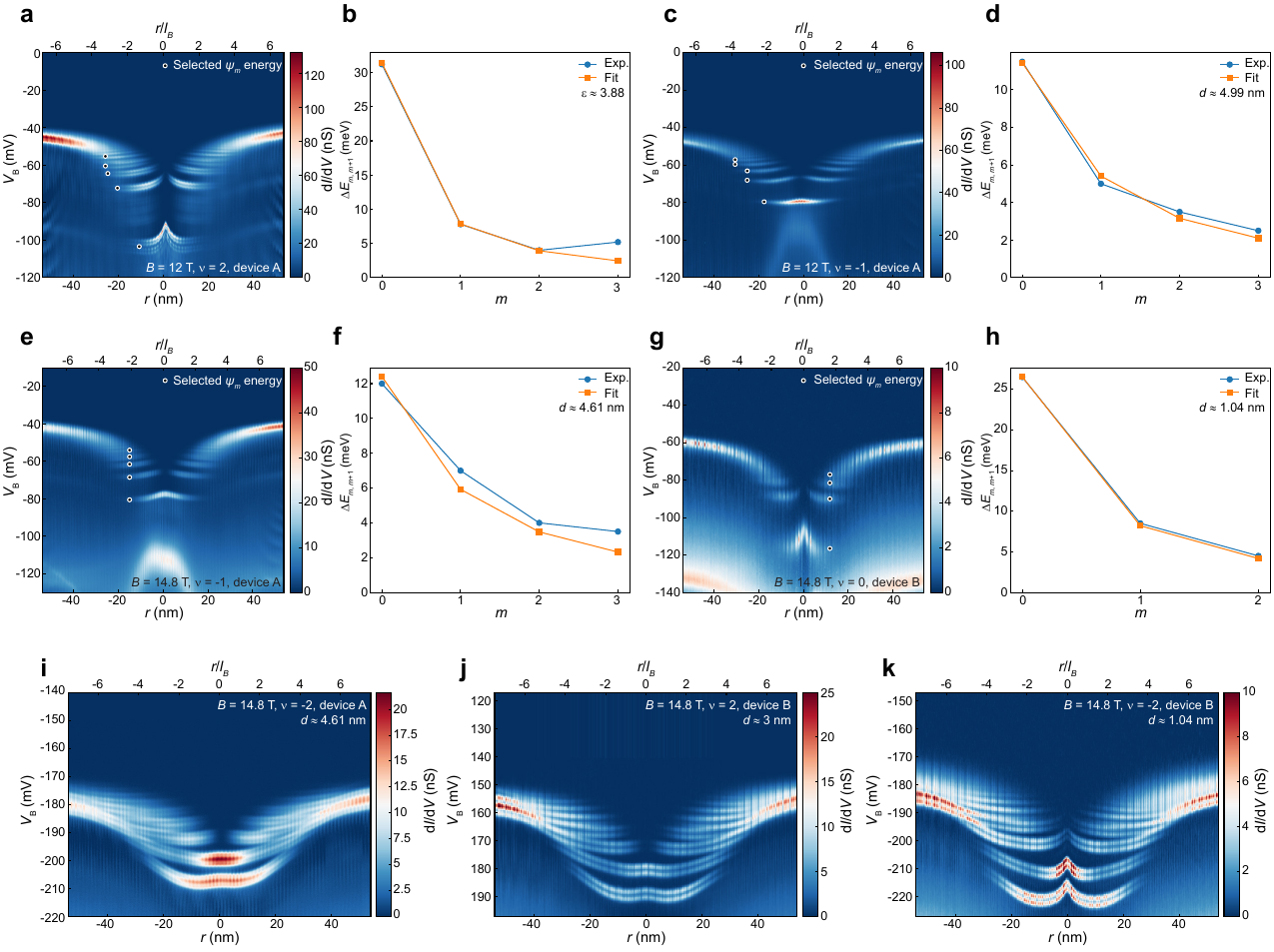}
\end{figure*}
\noindent\textbf{Fig. S14 \textbar{} Estimation of subsurface defect depths.} \textbf{a}, \textbf{b}, Determination of the dielectric constant using a surface $+e$ defect in device A at $B=12$ T and $\nu=2$. The energy splittings between adjacent Landau orbitals, $\Delta E_{m,m+1}$, are extracted from the bias-voltage separations between the corresponding spectral peaks marked in \textbf{a}. For larger $m$, the spectral features overlap, and only orbitals that can be reliably resolved are included. The measured $\Delta E_{m,m+1}$ are fitted using the perturbative model in Methods with $d=0$ (\textbf{b}), yielding a relative dielectric constant of $\epsilon\approx3.88$. In the following calculations, we approximate this value by $\epsilon=4$.

\textbf{c}-\textbf{h}, Depth determination for different subsurface $+e$ defects using the same procedure as in \textbf{a} and \textbf{b}. The fitted depths are $d\approx4.99$ nm for the defect shown in \textbf{c} and \textbf{d} (also Fig. 2c), $d\approx4.61$ nm for the defect shown in \textbf{e} and \textbf{f}, and $d\approx1.04$ nm for the defect shown in \textbf{g} and \textbf{h} (also Fig. 4e).

\textbf{i}-\textbf{k}, Depth estimation for the switchable subsurface $-e$ defect in device B. \textbf{j} shows the corresponding STS data acquired at $B=14.8$ T and $\nu=2$. Because the defect is charged only for $V_\text{G}>1.56$ V, it is not possible to measure the orbital splitting of the N = 0 LL within an integer quantum Hall gap. Instead, we rely on the orbital splitting pattern of the N = 1 LL. However, the N = 1 orbital splitting is strongly affected by tip effects (Sec. \ref{subsubsec:tip-sim}), preventing a reliable comparison with theoretical predictions. We therefore adopt an interpolation approach by comparing the measured N = 1 orbital splitting patterns with the N = -1 orbital splitting patterns of $+e$ charged defects at $\nu = -2$ whose depths are already known. The magnitude of the splitting for the unknown $-e$ defect lies between that of the $d\approx4.61$ nm defect shown in \textbf{i} and the $d\approx1.04$ nm defect shown in \textbf{k} (the bias voltage scale is the same from \textbf{i} to \textbf{k}). Based on this comparison, we estimate the depth of the defect shown in \textbf{j} (also Fig. S\ref{fig:gate_switch_defect}, and Fig. 4c) to be approximately $d\approx3$ nm. The spectrum in \textbf{j} is inverted for ease of comparison, exploiting the particle-hole symmetry of the system.

STM set conditions: (\textbf{a}) $V_\text{G}=1.065$ V, $V_\text{B}=-200$ mV, $I_\text{t}=2$ nA, $V_\text{rms}=0.5$ mV; 
(\textbf{c}) $V_\text{G}=-0.45$ V, $V_\text{B}=-200$ mV, $I_\text{t}=2$ nA, $V_\text{rms}=0.5$ mV;
(\textbf{e}) $V_\text{G}=-0.57$ V, $V_\text{B}=-170$ mV, $I_\text{t}=1$ nA, $V_\text{rms}=0.5$ mV; 
(\textbf{g}) $V_\text{G}=-0.045$ V, $V_\text{B}=-1$ V, $I_\text{t}=100$ pA, $\Delta z=-290$ pm, $V_\text{rms}=1$ mV; 
(\textbf{i}) $V_\text{G}=-1.25$ V, $V_\text{B}=-300$ mV, $I_\text{t}=1$ nA, $V_\text{rms}=0.5$ mV; 
(\textbf{j}) $V_\text{G}=2.2$ V, $\Delta z=-320$ pm, $V_\text{rms}=0.5$ mV, reference setpoint $V_\text{G}=1$ V, $V_\text{B}=-1$ V, $I_\text{t}=100$ pA;
(\textbf{k}) $V_\text{G}=-2.05$ V, $V_\text{B}=-1$ V, $I_\text{t}=100$ pA, $\Delta z=-280$ pm, $V_\text{rms}=1$ mV.
\clearpage


\end{document}